\documentclass[aps,onecolumn,preprintnumbers,nofootinbib,superscriptaddress]{revtex4}
\usepackage{amsmath} \usepackage{graphicx}  \usepackage{amsfonts}
\usepackage{array} \usepackage{amsthm} \usepackage{bm}
\usepackage{palatino} \usepackage{mathpazo} 
\usepackage{supertabular}
\usepackage{subfig}
\usepackage[breaklinks]{hyperref}
\usepackage{color}
\usepackage[font=small,labelfont=bf,
justification=justified,
format=plain]{caption} %
\usepackage{graphicx}
\usepackage{caption}
\usepackage[font=small,labelfont=bf,justification=justified]{caption}
\usepackage{color}

\usepackage{booktabs}
\usepackage{multirow}

\usepackage[labelfont=bf,labelsep=quad,justification=justified]{caption}

\usepackage{epstopdf}

\newcommand{\be}{\begin{equation}}
\newcommand{\ee}{\end{equation}}
\newcommand{\ba}{\begin{eqnarray}}
\newcommand{\ea}{\end{eqnarray}}
\newcommand{\bal}{\begin{align}}
\newcommand{\eal}{\end{align}}

\newcommand{\ro}{\rho}

\newcommand{\bw}{\begin{widetext}}
	\newcommand{\ew}{\end{widetext}}

\begin{document}

\title{Distinguishing rotating Kiselev black hole from naked singularity using spin precession of test gyroscope}

\author{Muhammad Rizwan}\email{m.rizwan@sns.nust.edu.pk}

\affiliation{Department of Mathematics, School of Natural
	Sciences (SNS), National University of Sciences and Technology
	(NUST), H-12, Islamabad, Pakistan}
	
\author{Mubasher Jamil}\email{mjamil@sns.nust.edu.pk}

\affiliation{Department of Mathematics, School of Natural
	Sciences (SNS), National University of Sciences and Technology
	(NUST), H-12, Islamabad, Pakistan}

\author{Anzhong Wang}\email{anzhong$\_$wang@baylor.edu}
\affiliation{GCAP-CASPER, Physics Department, Baylor University, Waco, TX 76798-7316, USA\\ 
Institute for Advanced Physics $\&$ Mathematics, Zhejiang University of Technology, Hangzhou, 310032, China}

\date{\today}

\begin{abstract}
We study the critical values of the quintessential and spin parameters, to distinguish a rotating Kiselev black hole (RKBH) from a naked singularity. For any value of the dimensionless quintessential parameter $\omega_{q} \in (-1, -1/3)$, when increasing the value of quintessential parameter $\alpha$, the size of the event horizon increases, 
whereas the size of the outer horizon decreases. We then study the spin precession of a test gyroscope attached to a stationary observer in this spacetime. Using the spin 
precessions we differentiate black holes from naked singularities. If the precession frequency becomes  large, as approaching 
to the central object in the quintessential field along any direction, then the spacetime is a black hole. A spacetime will contain a naked 
singularity if the precession frequency remains finite everywhere except at the singularity itself. Finally, we study the Lense-Thirring  
precession frequency for rotating Kiseleb black hole and the geodetic precession for  Kiselev black hole.
\end{abstract}

\maketitle



\section{Introduction}
Current research on observational measurements predicts the accelerating expansion of Universe, which is due to a presence of state with the negative pressure \cite{1,2,3}.  The negative pressure could be due to a cosmological constant or a so-called ``Dark Energy" \cite{QM1}. This energy is responsible for repulsive gravitational effects in the recent Universe and usually it is modal as exotic fluid. The fluid can be considered such that the state parameter $\omega$ is ratio of fluid pressure $p$ to its density $\ro$, that is $\omega=p/\ro$. The different models for dark energy have been proposed among: quintessence\cite{KBH,QU2}, phantom dark energy \cite{phantom1,phantom2}, quintom \cite{quintom1,quintom2}, K-essence \cite{ksc} and others. The difference of these models for dark energy is the value of parameter $\omega_{q}$. 

To study the dynamics of recent Universe we have to consider repulsive gravitational effects caused by negative pressure due to presence of dark energy. Quintessence is a candidate of dark energy according to which the dimensionless quintessential state parameter $\omega_{q}$ is related to the pressure 
$p$ and energy density $\rho$ of the quintessential field through the equation of state of the quintessential field, 
$p=\omega_{q} \rho$ \cite{KBH}. Furthermore, the range of the parameter $\omega_{q}$ is $-1<\omega_{q}<-1/3$ \cite{wq1,wq2,wq3}. If the quintessence matter exists all over the Universe, it can also be around a black hole. The spherically symmetric static black hole in a quintessential matter field which is generalization of Schwarzschild black hole and Schwarzschild AdS black hole is known as the Kiselev black hole (KBH) \cite{KBH}. The KBH and its charged version have been discussed in different aspects. Thermodynamics and phase transition of the charged KBH were studied in \cite{KBHTD,KBHTD2,KBHTD3}. The strong gravitational lensing by a KBH and a charged KBH has been discussed in \cite{lensing1,lensing2}. Recently, using Newman and Janis' technique \cite{NJanis} and its modification \cite{MAANJ},  rotational generalization of a KBH  have been given in \cite{QBH}. The Kerr-Newman-AdS black hole solution in a quintessential matter field has been also obtained in \cite{KerrADs}.

Due to the rotation of the central object, spacetime exhibits effects of the Lense-Thirring (LT) precession, which causes the dragging of locally inertial frames along the rotating spacetime \cite{LT,LTE,LT6}. Due to these effects a gyroscope attached to stationary observers in such a spacetime precesses with  certain frequencies. In the weak field approximations the magnitude of the precession frequency is proportional to the spin parameter of the central object and decreases with a cubic order of the distance from the central object \cite{LTE,LT6}. The gyroscope also precesses due to the spacetime curvature of the central object and this type of precessions is known as geodetic precessions or de Sitter precessions \cite{geodetic,RindlerBook}. These two effects are predicted by Einstein's theory of general relativity. To test these aspects of general relativity and to measure the precession rate due to the LT and geodetic effects relative to the Copernican system or the fixed star HR8703, known as IM Pegasi, of a test gyro due to the rotation of the Earth, Gravity Probe B has been launched  \cite{gprob}.The geodetic precession in the Schwarzschild black hole and the KBH have been studied in \cite{Geo,HartlelBook,Geodetic}. The LT-precession in the strong gravitational field of the Kerr and  Kerr-Taub-NUT black holes has been discussed in \cite{LTpr}.   

During the gravitational collapse of massive stars, the existence of naked singularities is the topic of great interest for researchers in the field of gravitational theory and relativistic astrophysics. The key question is that how one can differentiate whether the ultimate product in the life cycle of the compact object under the self-gravity collapse is naked singularity or black hole? Mathematically, a black hole is solution of the Einstein field equations (EFE). A stationary vacuum Kerr solution of EFE is characterized by two parameters, namely the mass $M$ and angular momentum $J$ of the central object. If the spin parameter $a$ (angular momentum per unit mass) satisfies the condition $M\geq a$, Kerr solution represent black hole and the Kerr singularity is contained in the event horizon. However,  if $M<a$ the event horizon disappears,  represents the naked singularity. Recently,  Chakraborty et al \cite{LTkerr,NS} gave the criteria based on the spin precession frequency of a test gyroscope attached to both static and stationary observers, to differentiate black holes from naked singularities. Using these criteria the Kerr black hole and naked singularities are discussed.  

The novelty of the present paper is to differentiate rotating black holes in a quintessential matter (rotating KBH) from a naked singularity. A stationary rotating Kiselev solution of Einstein field equation is characterized by four parameter, black hole mass $M$, spin parameter $a$, dimensionless quintessential parameter $\omega_{q}$ and quintessential parameter representing the intensity of the quintessence energy $\alpha$. In this paper we will given the critical values of spin parameter $a_c$ and quintessential parameter $\alpha_c$ to differentiate black hole from naked singularity. Note that, the analysis for $\omega_{q}=-2/3$ is already discussed \cite{QBH}. We have generalized the earlier work for general $\omega_{q}$ and study the critical values of the quintessential and spin parameters. Then, using the criteria of the spin precession of a gyro in a rotating black hole in a quintessential matter field (RKBH) we shall carry out the analysis in this paper. We will also study the effects of quintessential energy on the LT precession frequency for RKBH and geodetic precession of KBH.

The rest  of the paper is organized as follows.
In Section I we present a brief introduction of the field,  while in Section II the RKBHs are discussed an critical values of the quintessential and spin parameters are presented to differentiate black holes from naked singularities. In the end of this section critical values of the quintessential parameter for a KBH are given. The spin precession of a test gyroscope in a RKBH is discussed in Section III, form which we have obtained the LT-precession of a gyrsocope in the RKBH and geodetic precession  in the KBH. In Section IV,  using the key observations of spin precessions of test gyroscopes attached to stationary observers in a RKBH, we differentiate black holes from naked singularities. Section V is devoted for some concluding remarks. 

\section{Rotating Kiselev black holes \label{secGE}}

The line element of a rotating Kiselev black hole can be written as \cite{QBH}
\begin{eqnarray}\label{BH}
ds^{2} &=&-\left( 1-\frac{2Mr+\alpha r^{1-3\omega_{q} }}{\Sigma }\right) dt^{2}+\frac{%
	\Sigma }{\Delta }dr^{2}-2a\sin ^{2}\theta \left( \frac{2Mr+\alpha r^{1-3\omega_{q} }}{%
	\Sigma }\right) d\phi dt+\Sigma d\theta ^{2}  \notag \\
&&+\sin ^{2}\theta \left[ r^{2}+a^{2}+a^{2}\sin ^{2}\theta \left( \frac{%
	2Mr+\alpha r^{1-3\omega_{q} }}{\Sigma }\right) \right] d\phi ^{2},  \label{1}
\end{eqnarray}%
where%
\begin{eqnarray*}
	\Delta  &=&r^{2}-2Mr+a^{2}-\alpha r^{1-3\omega_{q} }, \quad
	\Sigma  =r^{2}+a^{2}\cos ^{2}\theta.
\end{eqnarray*}
{  The associated stress-energy tensor of the quintessential field takes the form of \cite{QBH} with the quantities  $(\epsilon,p_r,p_\theta,p_\phi)$ being given by,
\begin{equation}
	\epsilon=-p_r=\frac{\alpha(1-3\omega_{q})r^{2-3\omega_{q}}}{8\pi\Sigma^2},\text{\ \ \ \ \ \ \ \ } p_\theta=p_\phi=\frac{\alpha(-1+3\omega_{q})[2r^2+(2-3\omega_{q})\Sigma]}{16\pi\Sigma^2}.
\end{equation}
{  It should be noted that here $M$ does not represent the total mass (or total energy) of the spacetime. By evaluating the Komar integrals, we will get the total mass $M_T$  interior to the surface $r=r_0$, and the corresponding total angular momentum $J_T$ of the RKBH, which are related to the mass $M$ and angular momentum $J$ of the Kerr black hole via the relations}
\begin{equation}
M_{T}=M+\alpha r^{-1-3\omega_{q}}_{0}\left[ \frac{r_0}{2}-\frac{\left( r^{2}_{0}+a^{2}\right) \arctan
	\left( a/r_{0}\right) }{a}\right], \text{\ \ \ \ \ \ \ \ }
J_T=J+\alpha r^{-1-3\omega_{q}}_{0}\left[ ar_{0}+\frac{r^3_{0}}{2a}-\frac{(r^2_{0}+a^2)\arctan(a/r_{0})}{4a^2}\right].
\end{equation}
In the absence of the quintessential matter, $\alpha=0$, the line element and other quantities reduces to that of
the Kerr black hole. To differentiate  black holes from naked singularities,  in this section we express the black hole parameters and radial distance in units of the gravitational mass: $a/M\rightarrow a,$ 
$\alpha M^{-1-3\omega_{q} }\rightarrow \alpha,$ $r/M\rightarrow r$ and suppose $b=1-3\omega_{q}$. {   Note that the event horizon must be a null surface. Since the RKBH is a stationary
	spacetime, the normal to the stationary surface must be proportional to $\partial_\alpha r$
	and such a surface is null, if $g^{\alpha\beta}(\partial_\alpha r) (\partial_\beta r)=g^{rr} = 0$. Thus, the event horizons are the roots of}
\begin{equation}\label{HEq}
\Delta =r^{2}-2r+a^{2}-\alpha r^{b}=0.
\end{equation}
The locations of the ergospheres are determined by the roots of $r^{2}+a^{2}\cos
^{2}\theta -2r-\alpha r^b=0$. The horizons and ergospheres completely depend on the choice of the black hole parameters. For
example, if we choose $\omega_{q} =-2/3,$ then~\eqref{HEq} is a cubic equation and has three
different real, three same real, or one real and two complex roots, depending on the discriminant, $\delta$=$\left(
36-27\alpha a^{2}\right)\alpha a^{2}-4a^{2}-32\alpha+4$ being positive, zero or negative,
respectively. Henceforth, the line element~\eqref{BH}
represents correspondingly black hole, extremal black hole, and naked singularity.

Generalizing the method used for the case $\omega_{q}=-2/3$ in \cite{QBH}, we parameterize the spin parameter as a function of $r$ and $\alpha$,
\begin{equation}\label{aa}
a^{2}(\alpha,r) =\alpha r^{b}-r^{2}+2r.
\end{equation}
The spin parameter has extrema for $\alpha=\alpha_{e}$,  given by
\begin{equation}\label{alphe}
\alpha_{e}(r)=\frac{2(r-1)}{b r^{b-1}}. 
\end{equation}
The extrema of $\alpha_e$ (denoted by $\alpha_c$) is located at $r=({b-1})/({b-2})$. So,
 the critical value of the quintessential parameter ($\alpha_{c}$) and the corresponding spin parameter ($a_{c}$) is given by   
\begin{equation}\label{CVinb}
\alpha_{c}=\frac{2(b-2)^{b-2}}{b({b-1})^{b-1}}, \quad \quad a_{c}(\alpha_{c})=\frac{b-1}{\sqrt{b(b-2)}}.
\end{equation}
In terms of $\omega_{q}$, \eqref{CVinb} takes the form
\begin{equation}\label{CVinw}
\alpha_{c}=\frac{2(-1-3\omega_{q})^{-1-3\omega_{q}}}{(1-3\omega_{q})({-3\omega_{q}})^{-3\omega_{q}}}, \quad \quad a_{c}(\alpha_{c})=\frac{{-3\omega_{q}}}{\sqrt{9\omega^{2}_{q}-1}}.
\end{equation}
For $\omega_{q}=-2/3$, we get $\alpha_{c}=1/6$ and $a_{c}(1/6)=2/\sqrt{3}$~\cite{QBH}. The expression for the critical value of the quintessential parameter $\alpha_{c}$ given by Eq. (23)  in \cite{KerrADs} is incorrect, because with their expression for $\omega_{q}=-1/2$, the critical value is $\sqrt{2}/5$ ($\approx0.28284$). But, if we choose $\alpha=0.29>\sqrt{2}/5$ and $a=1.2$, the spacetime \eqref{BH} represents black holes with three horizons. On the other hand, with our expression \eqref{CVinw}, for $\omega_{q}=-1/2$, $\alpha_{c}$ is equal to $8/15\sqrt{3}$ ($\approx0.30792$) and for any value of $\alpha>\alpha_{c}$ there does not exist a spin parameter $a$ for which \eqref{BH} represents a black hole spacetime. Thus, the corrected critical value of the quintessential parameter is given by \eqref{CVinw}. The critical values $\alpha_{c}$ and $a_c$ versus $\omega_{q}$ are shown in FIG.~\ref{criticalValue}, which show that by increasing $\omega_{q}$, both $\alpha_{c}$ and $a_c$ increase. Further, when $\omega_{q}\rightarrow-1$, $\alpha_{c}\rightarrow2/27$, we find  $a_c(\alpha_{c})\rightarrow 3/2\sqrt{2}$,  that is, for small $\omega_{q}$ both $\alpha_{c}$ and $a_c$ are finite. On the other hand, when $\omega_{q}\rightarrow-1/3$, $\alpha_{c}\rightarrow1$ 
 we have $a_c(\alpha_{c})\rightarrow \infty$,  which means that in the presence of the quintessential field with $\omega_{q}\approx-1/3$ and $\alpha \approx1$, a highly spinning black hole is formed.       
\begin{figure}[!ht]
	\minipage{0.50\textwidth}
	\includegraphics[width=8.8cm,height=6.0cm]{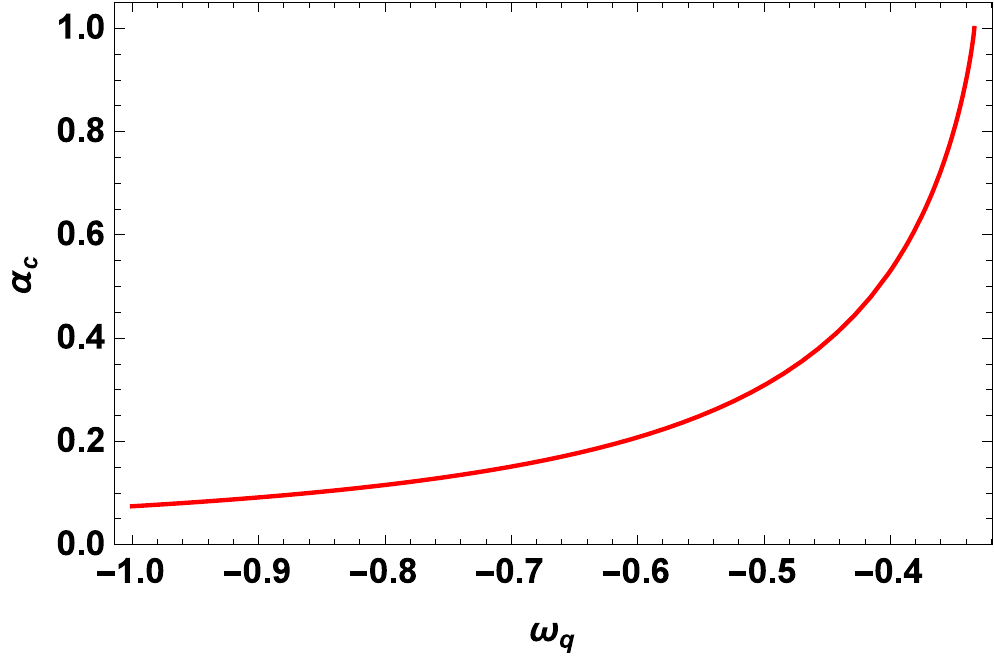}
	\label{7}
	\endminipage\hfill
	\minipage{0.50\textwidth}
	\includegraphics[width=8.8cm,height=6.0cm]{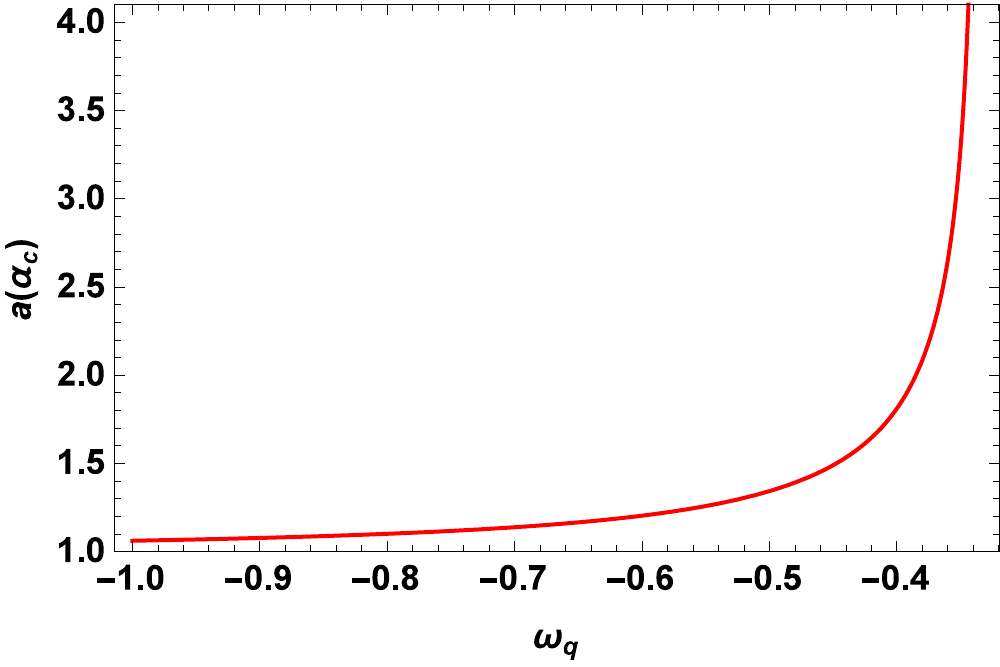}
	\label{6}
	\endminipage\hfill
	\caption{{\footnotesize The graph shows that with increasing $\omega_{q}$ the critical values of the quintessential  $\alpha_c$ and spin  $a_c$ parameters  increase. If $\omega_{q}\rightarrow-1$, we have $\alpha_{c}\rightarrow2/27$, $a_c(\alpha_{c})\rightarrow 3/2\sqrt{2}$,  and if $\omega_{q}\rightarrow-1/3$, we have $\alpha_{c}\rightarrow1$ and $a_c(\alpha_{c})\rightarrow \infty$.  }}
\label{criticalValue}
\end{figure}

\subsection{Black holes, extremal black holes and naked singularities}

In this subsection, we will discuss black holes, extremal black holes and naked singularities, represented by the line element~\eqref{BH}. The extrema of $\Delta$ can be obtained from the condition 
\begin{equation}\label{dHEwq}
\frac{d\Delta}{dr}=2(r-1)-\alpha(1-3\omega_{q})r^{-3\omega_{q}} =0.
\end{equation}
For any value of $\omega_{q}$, \eqref{dHEwq} has two real roots denoted by $r_{max}$ and $r_{min}$. Let us denote the corresponding extreme values of spin parameter $a$ by $a_{c}$ and $\overline{a}_{c}$, that is
\begin{equation}\label{acwq}
a_c=\sqrt{\alpha {r^{1-3\omega_{q}}_{min}}-{r^{2}_{min}}+2{r_{min}}}\quad and \quad \overline{a}_c=\sqrt{\alpha {r^{1-3\omega_{q}}_{max}}-{r^{2}_{max}}+2{r_{max}}}.
\end{equation}
Now we first develop our discussion for values of $\omega_{q}$ for which \eqref{dHEwq}, can be solve analytically and then we summarize the results for other values.
\subsubsection{$\omega_{q}=-1/2$}
For $\omega_{q}=-1/2$, ~\eqref{dHEwq} becomes
\begin{equation}\label{HEq1by2}
\frac{5}{2}\alpha r^{3/2}-2r+2=0.
\end{equation} 
and can be solve analytically. This equation has two real positive roots for $\alpha$ $\leq$ $\alpha_{c}={8}/{15 \sqrt{3}}$,  given by
\begin{equation}\label{rmin}
r_{min}=(\frac{4}{15 \alpha}+\sqrt{3} Im(u)-Re({u}))^{2}\quad \text{ and }\quad
r_{max}=(\frac{4}{15\alpha}+2Re({u}))^{2},
\end{equation}     
with
\begin{equation}\label{u}
u=\frac{1}{15\alpha }\left[ 2\left( 32-675\alpha ^{2}+{15\sqrt{3(675\alpha
		^{2}-64)}}\right) \right] ^{1/3},
\end{equation}
where $Re(u)$ and $Im(u)$ represent the real and imaginary part of $u$, respectively. These values play an important role in distinguishing  black holes from naked singularities. Note that for any value of $\alpha$ $\leq$ $\alpha_{c}={8}/{15 \sqrt{3}}$, there are three real positive roots of \eqref{HEq1by2}, $r_{-}$,~$r_{+}$ and $r_q$, representing, respectively,  the inner, event and outer horizons of the black hole. These horizons and extrema of $\Delta$ given by \eqref{rmin} satisfy the relation $r_{-} \leq r_{min}\leq r_{+}\leq r_{max}\leq r_{q}$. From FIG.~\ref{figrm1by2} (a), one can see that for small $\alpha$, $r_{max}$ becomes  very large, hence we conclude that the quintessential horizon $r_q$ is very large while with increasing $\alpha$ its size decreases. On the other hand, with increasing $\alpha$, $r_{min}$ increases and hence the size of the event horizon increases,  although  very slowly. Further,  if one of the equality holds,  that is, either $r_{min}$ or $r_{max}$ becomes the horizon of the black hole, \eqref{BH} represents an extremal black hole.

\begin{figure}[!ht]
	\minipage{0.50\textwidth}
	\includegraphics[width=8.8cm,height=6.0cm]{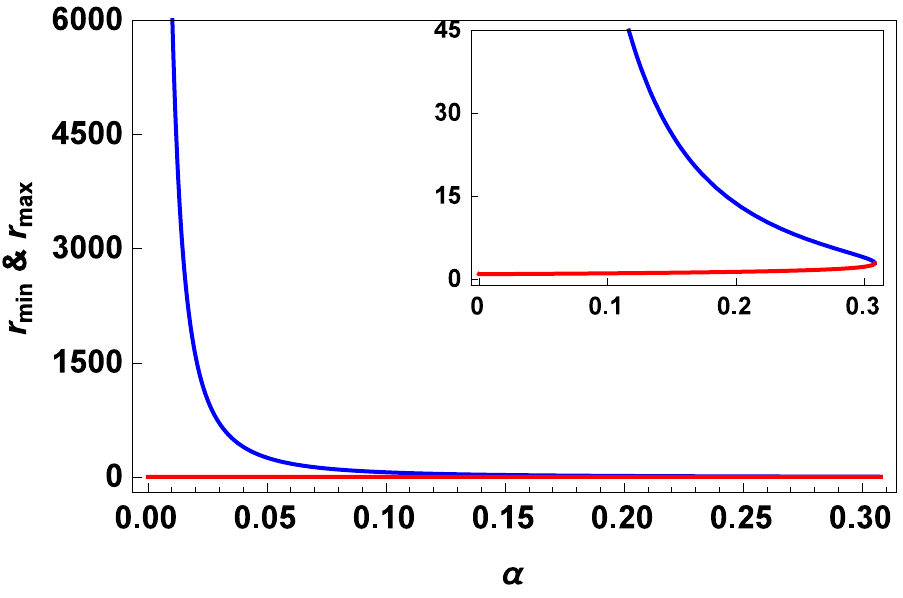}\\~~~~~~~~~~~~~~~~(a) 
	\label{6}
	\endminipage\hfill
	\minipage{0.50\textwidth}
	\includegraphics[width=8.8cm,height=6.0cm]{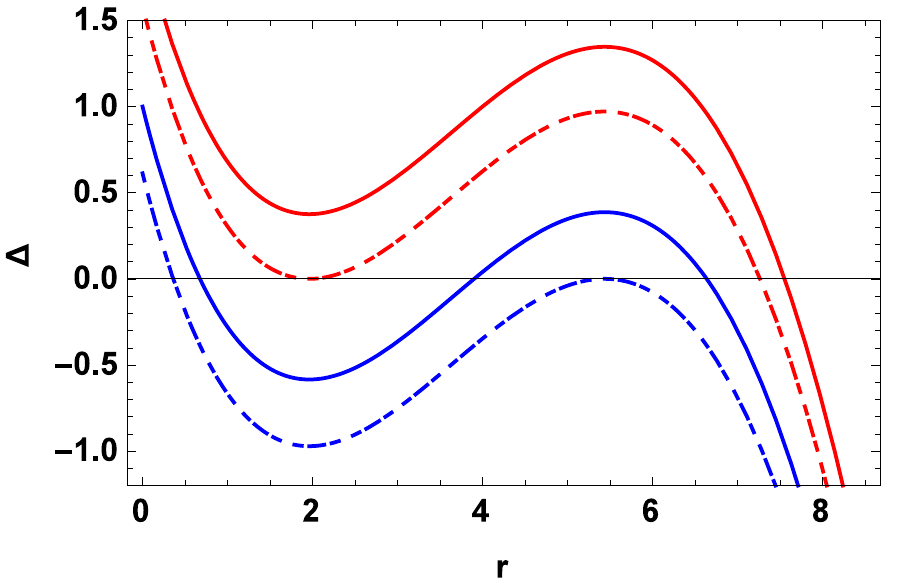}\\~~~~~~~~~~~~~~(b) 
	\label{7}
	\endminipage\hfill
	\caption{{\footnotesize (a): The graph plotted for $r_{min}$ (red curve) and $r_{max}$ (blue curve) against the parameter $\alpha$, which shows $r_{max}$ decrease when increasing the value of $\alpha$, while $r_{min}$ increases when increasing $\alpha$. The inset shows the variation in small scales, from which we can see that  $r_{min}$ and $r_{max}$ coincide for $\alpha_{c}={8}/{15 \sqrt{3}}$. (b): In this graph we have plotted $\Delta$ against $r$ for $\alpha=0.29$ and different values of $a$. The blue curve (plotted for $a=0.9$) shows a black hole with three horizons $r_{-}, r_{+}$ and $r_{q}$. The extreme black hole of Type1  (with $a=1.258693$), when $r_{-}=r_{+}$, is represented by the dashed red curve. The extreme black hole of Type2 (with $a=0.782936$), when $r_{+}=r_{q}$, is represented by the dashed blue curve. The red curve represents a naked singularity (with $a=1.3$) with one horizon $r_{q}$ only.}}\label{figrm1by2}
\end{figure}
\begin{figure}[!ht]
	\minipage{0.50\textwidth}
	\includegraphics[width=8.8cm,height=6.2cm]{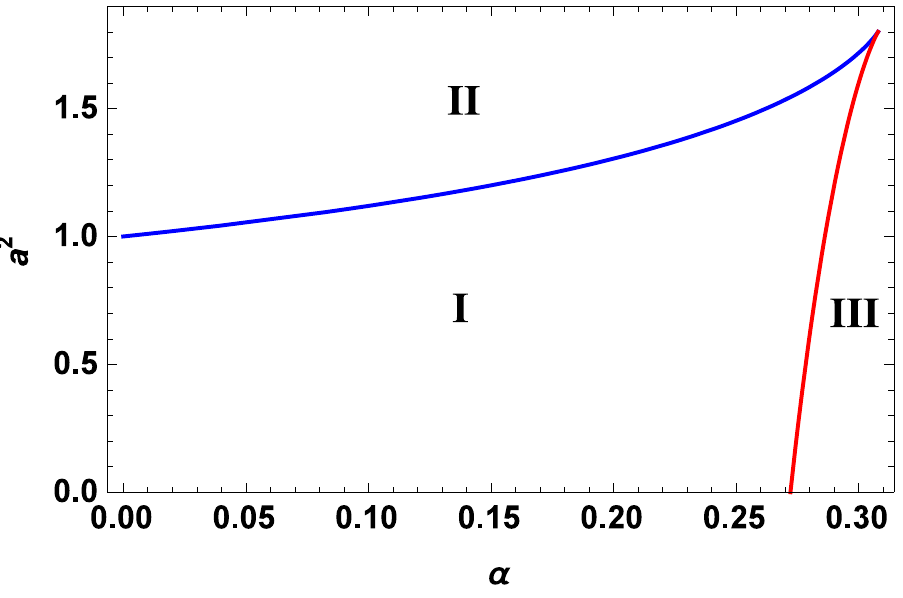} 
	\label{7}
	\endminipage\hfill
	\caption{{\footnotesize \emph{Region I} represents black holes with three horizons. The boundary of  Regions I and II represents extremal black holes of Type1 and the boundary of Regions I and III represents extremal black holes of Type2. From the figure we can see that $\overline{a}_{c}$ is defined only for $0.272166<\alpha<0.307920$. Thus,  \eqref{BH} represents extremal black holes of Type2 only for these values of $\alpha$. For the values of $\alpha$ and $a$ at the point of intersection of these curves, \eqref{BH} represents super extremal black holes. For all values of $\alpha$ and $a^2$ in Regions II and III,  \eqref{BH} represents naked singularities}}\label{figrg1by2}
\end{figure}

The extremal black hole can be of three types.

\textbf{\emph{Type1:}}
The first type of the extremal black hole exists when $r_{min}$ is the horizon of the black hole, for which $\Delta(r_{min})=0$. Here, the inner and event horizons merge into a single horizon,  that is,  $r_{-}=r_{+}$~\footnote{In \cite{QBH} [cf. Eq.(40)], for the case of $\omega_{q}=-2/3$, it is claimed that the other extrema of $\Delta$ ($r_{max}$ in our case) is the outer horizon of the black hole. But this is not true in general and $r_{max}<r_{q}$ as shown in FIG.\ref{figrm1by2} (b).}. For this type of black holes the spin parameter satisfies the condition  
\begin{equation}\label{ac1by2}
a_c=\sqrt{\alpha {r^{5/2}_{min}}-{r^{2}_{min}}+2{r_{min}}}~,
\end{equation}
 where $r_{min}$ is given by~\eqref{rmin}. This case is shown by the red dashed curve in FIG.\ref{figrm1by2} (b).
   
 \textbf{\emph{Type2:}}
 The second type of the extremal black hole exists when $r_{max}$ is the horizon of the black hole, where $\Delta(r_{max})=0$. In this case, the event and outer horizons merge into a single horizon, that is, $r_{+}=r_{q}$. This type of the extremal black holes is formed when the rotation parameter satisfies the condition
 \begin{equation}\label{ac1by2bar}
 \overline{a}_c= \sqrt{\alpha {r^{5/2}_{max}}-{r^{2}_{max}}+2{r_{max}}}~,
 \end{equation}
 where $r_{max}$ is given by~\eqref{rmin}. This case is shown by the dashed blue curve in FIG.\ref{figrm1by2} (b). In addition, in this case $r_{-}<r_{min}$. Note that, the value of $\overline{a}_c$ is defined only for $0.27216<\alpha<0.30792$ (This can be see from the red curve in FIG.\ref{figrg1by2}). So, this type of extremal black holes exists only for these values of $\alpha$.  
 
 \textbf{\emph{Type3:}}
The third type of the extremal black hole exists when all the three horizons merge into a single horizon. In this case, we have  $r_{min}$~=~$r_{max}$, which is possible for 
 \begin{equation}\label{crtcl1by2}
\alpha=\alpha_{c}=\frac{8}{15 \sqrt{3}}\quad \text{ and }\quad
 a_{c}=\overline{a}_c=\frac{3}{\sqrt{5}}.
 \end{equation}
This type of black hole is known as super-extremal black holes. 

Finally, we conclude that for any value of $\alpha<\alpha_{c}$ and the corresponding spin parameter $a<{a}_c$,  \eqref{BH} can represents a black hole with three different horizons. For any given value of $\alpha$ with $\alpha < \alpha_{c}$, the line element \eqref{BH} represents extremal black holes of \emph{Type1} or \emph{Type2}, depending on whether the spin parameter $a=a_c(\alpha)$ or $a=\overline{a}_c(\alpha)$. For $\alpha=\alpha_c$ and $a=a_c(\alpha_c)$, the spacetimes  of Eq.\eqref{BH} represent super-extremal black holes. For any other possibilities, the spacetimes represent naked singularities. In FIG.\ref{figrg1by2}, we have plotted $a=a_c$ (the blue curve) and $a=\overline{a}_c$ (the red curve)  in the $(\alpha, a^2)$-plane, which divide the whole plane into three different regions. In Region I, \eqref{BH} represents black holes,  whereas in Regions II and III it represents naked singularities. For all points on the boundary of Regions I and II, \eqref{BH} represents extremal black holes of \emph{Type1}, and on the boundary of Regions II and III \eqref{BH} represents extremal black holes of \emph{Type2}.

\subsubsection{When $\omega_{q}=-4/9$}

 Again for $\omega_{q}=-4/9$, \eqref{dHEwq} can be solve analytically. In this case  $\alpha_{c} \approx 0.404975$ and for all $\alpha\leq\alpha_c$ \eqref{dHEwq} have two real positive roots 
\begin{equation}\label{rm4by9}
r_{\min }=\left(\frac{3}{14 \alpha }+\frac{v}{2}-\frac{1}{2} \sqrt{\frac{27}{49 \alpha ^2}-v^2+\frac{54}{343 \alpha ^3 v}}\right)^3,
\quad \quad
r_{max}=\left(\frac{3}{14 \alpha }+\frac{v}{2}+\frac{1}{2} \sqrt{\frac{27}{49 \alpha ^2}-v^2+\frac{54}{343 \alpha ^3 v}}\right)^3,
\end{equation}  
with
\begin{equation}\label{v}
v=\sqrt{\frac{9}{49 \alpha ^2}+\frac{2^{2/3} ({27+\sqrt{729-10976 \alpha ^3}})^{1/3}}{7 \alpha }+\frac{2^{7/3}}{({27+\sqrt{729-10976 \alpha ^3}})^{1/3}}}.
\end{equation}
Corresponding to these two solutions we obtain the extreme values of the spin parameter 
\begin{equation}\label{ac4by9}
a_c= \sqrt{\alpha {r^{7/3}_{min}}-{r^{2}_{min}}+2{r_{min}}}~\quad and \quad  \overline{a}_c= \sqrt{\alpha {r^{7/3}_{max}}-{r^{2}_{max}}+2{r_{max}}}~.
\end{equation} 
where $r_{min}$ and $r_{max}$ are given by \eqref{rm4by9}. From FIG.\ref{fig4by9} (a), we can see that by increasing $\alpha$, $r_{min}$ increases and $r_{max}$ decreases. Thus, due to the relation $r_{-} \leq r_{min}\leq r_{+}\leq r_{max}\leq r_{q}$, in this case we can also conclude that by increasing $\alpha$, the size of the event horizon increases, while that of the outer horizon decreases. For $\alpha<\alpha_c$ and $a<a_{c}$,~\eqref{BH} represents black holes with three horizons (as shown by the blue curve in FIG.\ref{fig4by9} (b)). For $\alpha<\alpha_c$, and $a=a_c$ or $a=\overline{a}_c$, \eqref{BH} represents extramal black holes of \emph{Type1} or \emph{Type2} (as shown by the dashed red and the dashed blue curve in FIG.\ref{fig4by9} (b)). Further, FIG.\ref{figrg4by9} shows that extremal black holes of \emph{Type2} exist only for $0.375<\alpha<0.404975$. For any other possibilities (that is for any $(\alpha, a^2)$ in Regions II or III of FIG.\ref{figrg4by9}), the line element \eqref{BH} represents naked singularities.
\begin{table}[]
	\centering
	
	\begin{tabular}{|c|c|c|c|c|c|c|}
		\hline
	
	$\boldsymbol{\omega_{q}}$                                                                              & -4/9 &-1/2& -5/9 & -2/3  & -7/9  & -8/9  \\ \hline
	$\boldsymbol{\alpha_{c}}$                                                                              & 0.404975 & 0.30792 & 0.244298 & 0.166667 & 0.121934 & 0.0934523 \\ \hline

  $\boldsymbol{\overline{\alpha}_{c}}$                                                                 & 0.375 & 0.2721655269759& 0.2051971136011 & 0.125 & 0.0806490313479 & 0.053963051556 \\ \hline
 	 $\boldsymbol{a(\alpha_{c})}$                                                                      & 1.51186 & 1.34164 & 1.25 & 1.1547 & 1.1068 &1.07872 \\ \hline 
	\end{tabular}
\caption{Critical values quintessential parameter $\alpha$ and spin parameter $a$ for different $\omega_{q}$.}\label{table1}
\end{table}

\begin{table}[]
	\centering
	\begin{tabular}{|c|c|c|c|c|}
		\hline
		\textbf{Type} & \textbf{Horizons} & \textbf{Condition} & \textbf{ Range of } $\boldsymbol{\alpha}$ & \textbf{Condition of a} \\ \hline
	1 & $r_{-}=r_{+}$        & $\Delta(r_{min})=0$         & $\alpha<\alpha_c$       & $a=a_c$                   \\ \hline
	2	& $r_{+}=r_{q}$        & $\Delta(r_{max})=0$         & $\overline{\alpha}_c<\alpha<\alpha_{c}$         & $a=\overline{a}_c$          \\ \hline
	3& $r_{-}=r_{+}=r_{q}$        & $r_{min}=r_{max}$         & $\alpha=\alpha_{c}$        & $a=a(\alpha_{c})$          \\ \hline
	\end{tabular}
\caption{{\footnotesize We are given the conditions for different types of extremal black holes. }}\label{table2}
\end{table}

 By numerical analysis we see that the behavior is same for all value of $-1<\omega_{q}<-1/3$, that is, \eqref{BH} represents black holes, provided that  $\alpha\leq\alpha_c$ and $a\leq a_{c}$. By increasing $\alpha$, the size of the quintessential horizon decreases, while that of event horizon increases. Further, the types of extremal black hole and conditions for some values of $\omega_{q}$ are summarized by the Tables \ref{table1} and \ref{table2} \footnote{The values $\overline{\alpha}_{c}$ in Table \ref{table1} also plays an important role in Kiselev black hole. For any chosen $\omega_{q}$, $\alpha=$$\overline{\alpha}_{c}$ is critical value for Kiselev black hole. The line element for Kiselev represent black hole only for $\alpha\leq$$\overline{\alpha}_{c}$.}.
\begin{figure}[!ht]
	\minipage{0.50\textwidth}
	\includegraphics[width=8.8cm,height=6.0cm]{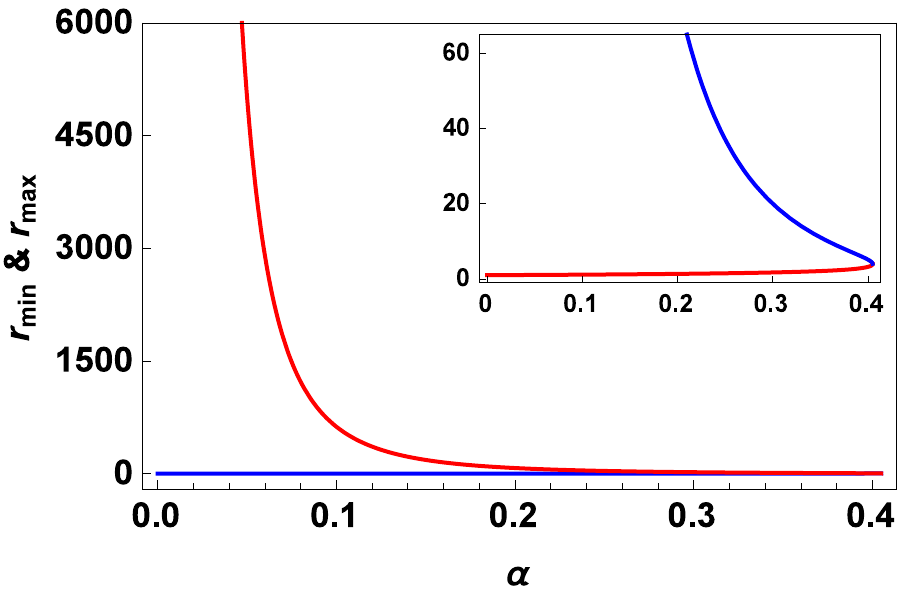}\\~~~~~~~~~~~~~(a) 
	\label{6}
	\endminipage\hfill
	\minipage{0.50\textwidth}
	\includegraphics[width=8.8cm,height=6.0cm]{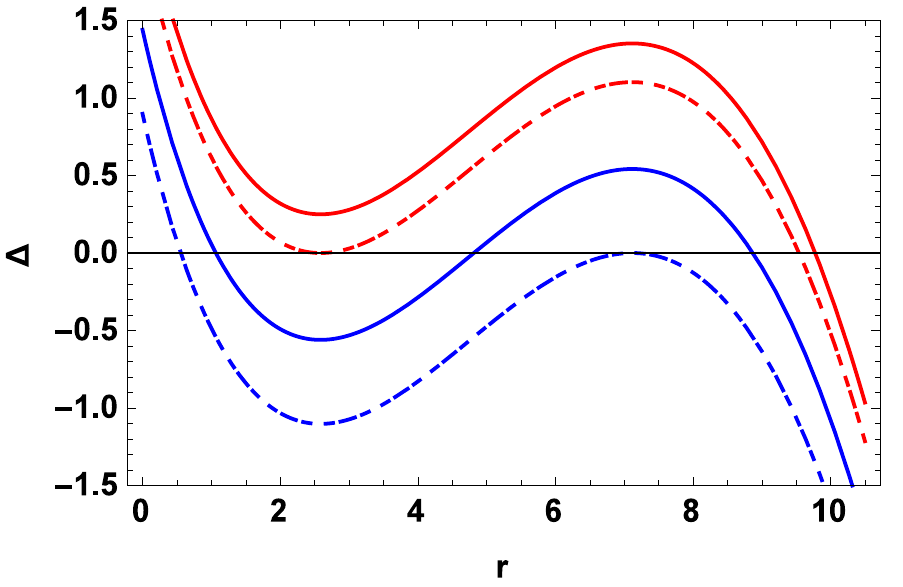}\\~~~~~~~~~~~~~~(b) 
	\label{7}
	\endminipage\hfill
	\caption{{\footnotesize (a): The graph plotted for $r_{min}$ (red curve) and $r_{max}$ (blue curve) against parameter $\alpha$ which shows $r_{max}$ decreases with increasing the value of $\alpha$ while $r_{min}$ increases with increasing $\alpha$. The inset shows variation in small scales from which we can see that $r_{min}$ and $r_{max}$ coincide for $\alpha_{c}=0.404975$. (b): In this graph we have plotted $\Delta$ against $r$ for $\alpha=0.383$ and different values of $a$. The blue curve (plotted for $\alpha=1.2$) shows black hole with three horizons $r_{-}, r_{+}$ and $r_{q}$. The dashed red curve (plotted for $a=1.41428$) represents extreme black hole of Type1 when $r_{-}=r_{+}$. The dashed blue curve (plotted for $a=0.947639$) represents extreme black hole of Type2 when $r_{+}=r_{q}$. The red curve (plotted for $a=1.5$) represent naked singularity with one horizon $r_{q}$ only.}}\label{fig4by9}
\end{figure}
\begin{figure}[!ht]
	\centering
	\minipage{0.50\textwidth}
	\includegraphics[width=8.8cm,height=6.2cm]{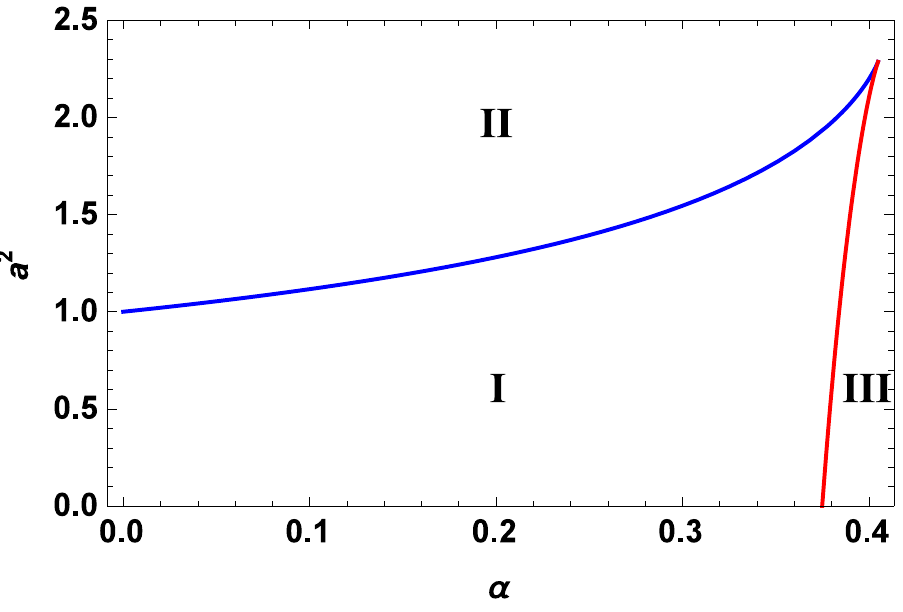} 
	\label{7}
	\endminipage\hfill
	\caption{{\footnotesize \emph{Region I} represent black hole with three horizons. Boundary of region I and II represent extremal black hole of \emph{Type1} and boundary of region I and II represent extremal black hole of \emph{Type2}. The figure shows that the extremal black hole of Type2 exist only for $0.375<\alpha<0.404975$. For the values in region II and III spacetime \eqref{BH} represent naked singularities.}}\label{figrg4by9}
\end{figure}

\section{Spin Precession of test gyroscope in RKBH}
In this section, we will discuss the spin precession frequency of a test gyroscope attached to a stationary observer in a RKBH. {  A stationary observer is an observer who remains at fixed $r$ and $\theta$ coordinates by rotating around the black hole (with respect to observers at infinity) in the same sense as the black hole's rotation. The 4-velocity of such an observer is $[u^\mu]=u^t(1,0,0,\Omega)$, where $\Omega=d\phi/dt$.} 
Consider a test gyroscope attached to stationary observer moving along Killing trajectory in a RKBH (stationary spacetime). The RKBH spacetime admit two Killing vectors: the time translation Killing vector $\partial_t$ and the azimuthal Killing vector $\partial_\phi$. The vector $K=\partial_t+\Omega\partial_\phi$ is also a Killing vector. The one-form of the general spin precession frequency can be expressed as \cite{Straumann}
\begin{equation}
\tilde{\Omega}_p=\frac{1}{2K^2}*(\tilde{K}\wedge d\tilde{K})
\end{equation}
 where  $\tilde{K}$ is corresponding covector of $K$, * represents the Hodge dual and $\wedge$ is wedge product. 
 Thus, spin precession frequency of a timelike stationary observer, having 
an angular velocity $\Omega$ with respect to a  fixed star in a stationary axisymmetric spacetime,  is given by \cite{NS}
\begin{equation}\label{SPD}
\vec{\Omega}_{p}=\frac{\varepsilon_{ckl}{ }}{2\sqrt{-g}%
	\left( 1+2\Omega \frac{g_{0c}}{g_{00}}+\Omega ^{2}\frac{g_{cc}}{g_{00}}\right) }%
\left[ \left( g_{0c,k}-\frac{g_{0c}}{g_{00}}g_{00,k}\right) +\Omega \left(
g_{cc,k}-\frac{g_{cc}}{g_{00}}g_{00,k}\right) +\Omega ^{2}\left( \frac{g_{0c}%
}{g_{00}}g_{cc,k}-\frac{g_{cc}}{g_{00}}g_{0c,k}\right) \right]\partial_l,
\end{equation}
where $g$ is the determinant of the metric $g_{\mu\nu}$ and $\varepsilon _{ckl}$ is the Levi-Civita symbol. This expression is valid for observers both inside and outside of the ergosphere for a restricted range of $\Omega$,  such that its velocity $u^\nu=u^t (1,0,0,\Omega)$, remains timelike. Substituting the metric coefficients from~\eqref{BH} into~\eqref{SPD}, we get
\begin{equation}\label{SP}
\vec{\Omega}_{p}=\frac{(F\sqrt{\Delta }\cos \theta) \hat{r}+(H\sin \theta )\hat{%
		\theta}}{\Sigma ^{3/2}\left[ \Sigma -\left( 2Mr+\alpha r^{b}\right) +2\Omega
	a\sin ^{2}\theta \left( 2Mr+\alpha r^{b}\right) -\Omega ^{2}\sin ^{2}\theta
	\left\{ \left( r^{2}+a^{2}\right) \Sigma +a^{2}\sin ^{2}\theta \left(
	2Mr+\alpha r^{b}\right) \right\} \right] },
\end{equation}%
with%
\begin{eqnarray} \label{F&H}
F &=&a\left( 2Mr+\alpha r^{b}\right) -\frac{\Omega }{8}\left[
8r^{4}+3a^{4}+8a^{2}r^{2}+16a^{2}Mr+4a^{2}\left( 2\Delta -a^{2}\right) \cos
2\theta +a^{4}\cos 4\theta \right]   \notag \\
&&+\Omega ^{2}a^{3}\sin ^{4}\theta (2Mr+\alpha r^{b}),  \notag \\
H &=&a\left[ M\left( r^{2}-a^{2}\cos ^{2}\theta \right) +\frac{\alpha}{2}%
r^{b-1}\left\{ (2-b)r^{2}-ba^{2}\cos ^{2}\theta \right\} \right]   \notag \\
&&+\Omega \left[ r\left( r^{4}+a^{4}\cos ^{4}\theta +2r^{2}a^{2}\cos
^{2}\theta \right) -r\left( r^{2}+a^{2}\cos ^{2}\theta \right) \left(
2Mr+\alpha r^{b}\right) +\left( r^{2}+a^{2}+a^{2}\sin ^{2}\theta \right) \times
\right.   \notag \\
&&\left. \left\{ -Mr^{2}+\left( M+\frac{\alpha}{2}br^{b-1}\right) a^{2}\cos
^{2}\theta +\frac{\alpha}{2}\left( b-2\right) r^{b+1}\right\} \right] +a\Omega
^{2}\sin ^{2}\theta \left[ r^{3}\left( 2Mr+\alpha r^{b}\right) +ra^{2}\cos
^{2}\theta \left( 2Mr+\alpha r^{b}\right) \right.   \notag \\
&&\left. +Mr^{4}+Mr^{2}a^{2}-(r^{2}+a^{2})\left( M+\frac{\alpha}{2}%
br^{b-1}\right) a^{2}\cos ^{2}\theta -\frac{\alpha}{2}(r^{2}+a^{2})\left(
b-2\right) r^{b+1}\right],
\end{eqnarray}%
where $b=1-3\omega_{q}$ and $\hat{r}$, $\hat{\theta}$ are the basis vectors in the $r$ and $\theta$ directions, respectively. Setting $\alpha=0$, gives the spin precession for a Kerr black hole \cite{NS}. In the above expression for timelike observers, $\Omega$ has the restriction 
\begin{equation}\label{omegarange}
\Omega _{-}(r,\theta )<\Omega (r,\theta )<\Omega _{+}(r,\theta ),
\end{equation}
where
\begin{equation}\label{AVm+-}
\Omega _{\pm }=\frac{a\sin \theta (2M\alpha+\alpha r^{b})\pm \Sigma \sqrt{%
		\Delta }}{\sin \theta \left[ \left( r^{2}+a^{2}\right) \Sigma +a^{2}\sin
	^{2}\theta \left( 2Mr+\alpha r^{b}\right) \right] }.
\end{equation}%
This expression shows that for an observer close to the horizon as well as to the ring singularity $(r=0, \theta=\pi/2)$, $\Omega_+$ and $\Omega_-$ coincide. Thus, no timelike observer can exist at these points and the expression for $\vec{\Omega}_{p}$ is not a valid expression at these points, but  still useful when discussing precession in the limit of these points. 

\subsection{LT-Precession Frequency}
The precession frequency ($\vec{\Omega}_{p}$) given by \eqref{SP} is the general precession frequency of gyroscope having angular velocity $\Omega$. The precession frequency include effects both due to spacetime rotation (LT-precession) and curvature (geodetic precession). If we set $\Omega=0$, then $\vec{\Omega}_{p}$ reduces to the LT-precession ($\Omega_{LT}$) frequency of the gyroscope attached to a static observer, who can exist only outside the ergosphere. The expression for the LT-precession frequency is
\begin{equation}\label{LT}
\vec{\Omega} _{LT}=a\frac{\left[ \left( 2Mr+\alpha r^{b}\right) \sqrt{\Delta }\cos \theta %
	\right] \hat{r}+\left[ \left\{ M\left( r^{2}-a^{2}\cos ^{2}\theta \right) +\frac{%
		\alpha }{2}r^{b-1}\left( \left( 2-b\right) r^{2}-ba^{2}\cos ^{2}\theta
	\right) \right\} \sin \theta \right] \hat{\theta} }{\left( r^{2}+a^{2}\cos
	^{2}\theta \right) ^{3/2}\left( r^{2}-2Mr+a^{2}\cos ^{2}\theta -\alpha
	r^{b}\right) }.
\end{equation}%
 The vector field of the LT-~precession frequency \eqref{LT}, for black holes and naked singularities for different values of $\alpha$ (in the Cartesian plane corresponding to $(r,\theta)$),  is plotted in the first and second row of FIG.\ref{VF}, respectively. It can be seen that for black holes, the LT- precession frequency diverges if the observer approaches  the ergosphere along any directions. However, outside the ergosphere the frequency is finite everywhere.   For naked singularities it is regular throughout the whole region except at the ring singularity ($r=0,~\theta=\pi/2$). This is because the denominator of \eqref{LT} goes to zero at the ergospheres and ring singularity. Further, for black holes the field lines   at the pole precess in the same direction as the black hole rotation, while those on the equatorial plane precess in the opposite sense as in the case of the linearized gravitation field \cite{RindlerBook}. 
\begin{figure}[!ht]
	\centering
	\minipage{0.30\textwidth}
	\includegraphics[width=2.0in,height=2.0in]{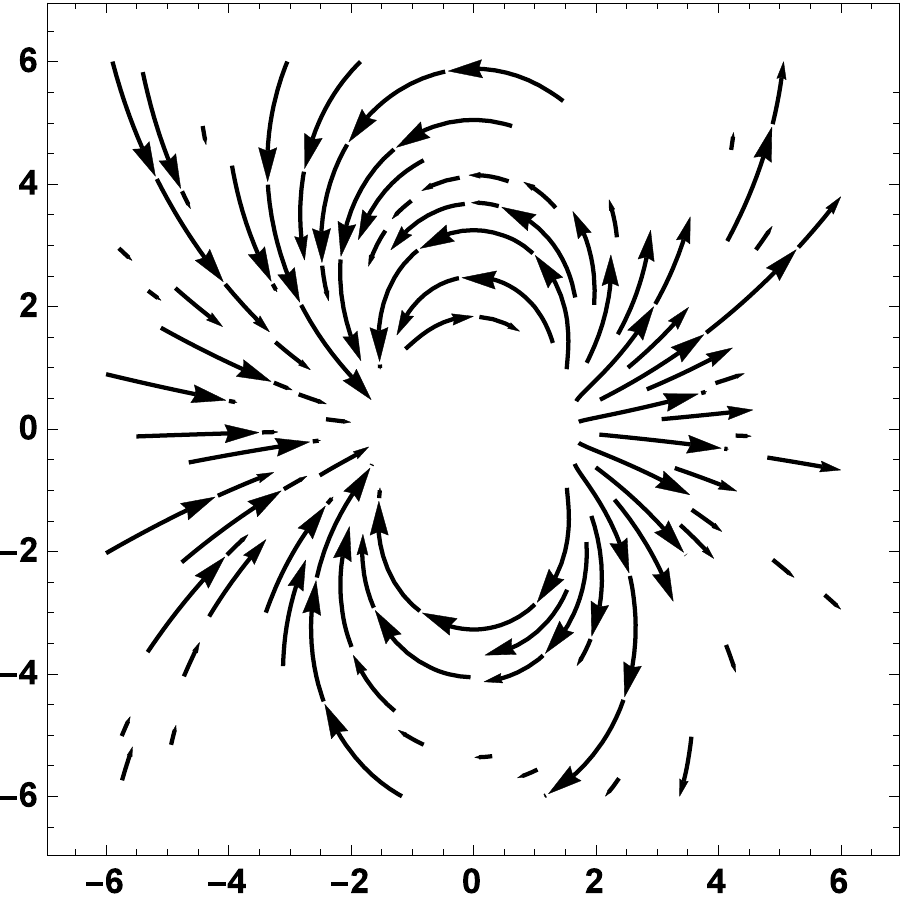}\\ (a) $\alpha=0$, $a=0.7$
	\label{7}
	\endminipage\hfill
	\minipage{0.30\textwidth}
	\includegraphics[width=2.0in,height=2.0in]{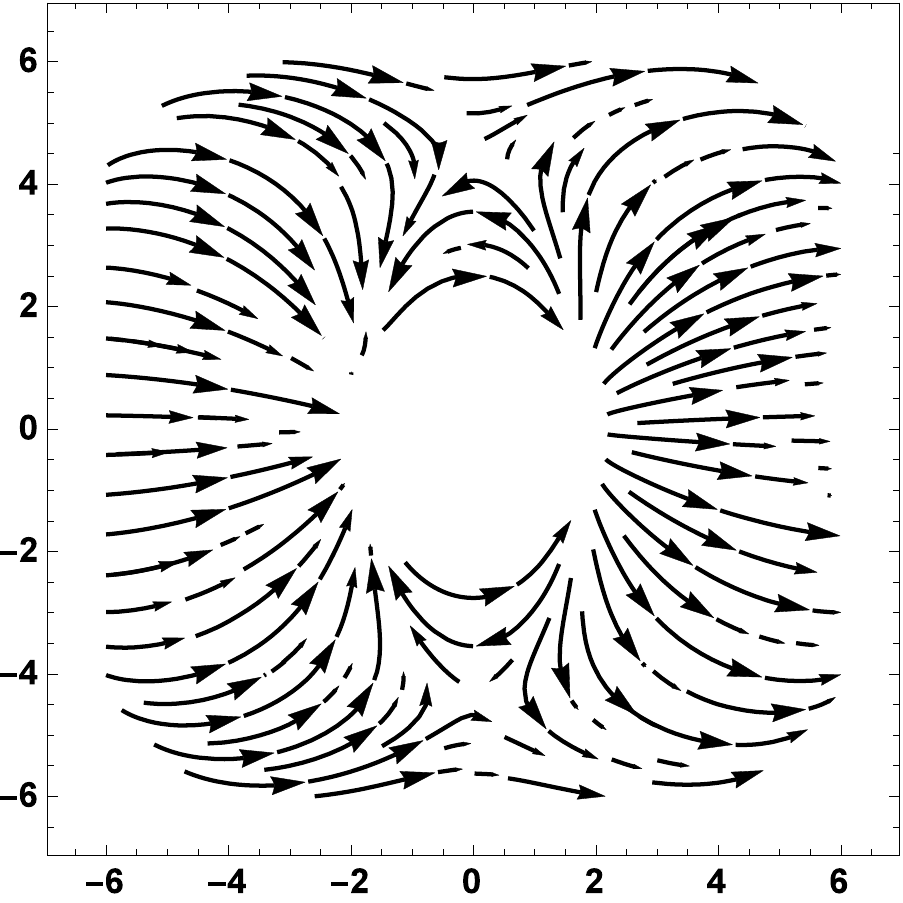}\\(b) $\alpha=0.1$, $a=0.8$, $\omega_{q}=-2/3$
	\label{7}
	\endminipage\hfill
		\minipage{0.30\textwidth}
		\includegraphics[width=2.0in,height=2.0in]{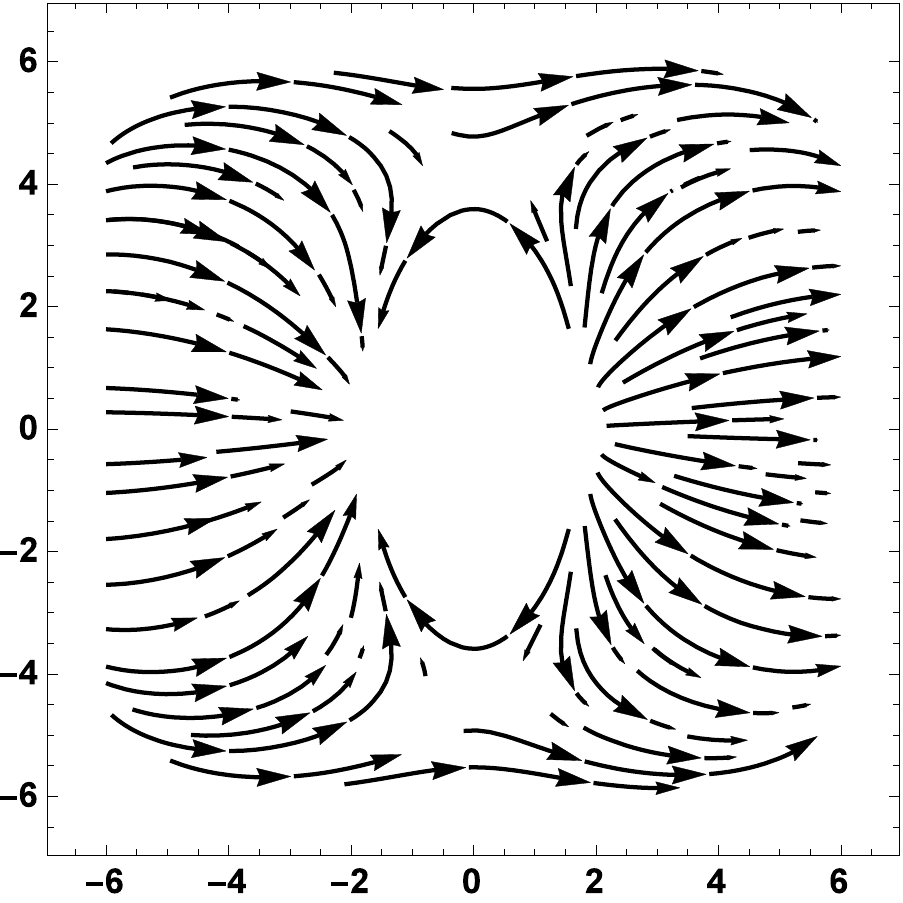}\\(c) $\alpha=0.05$, $a=0.6$, $\omega_{q}=-7/9$
		\endminipage\hfill
	\minipage{0.30\textwidth}
	\includegraphics[width=2.0in,height=2.0in]{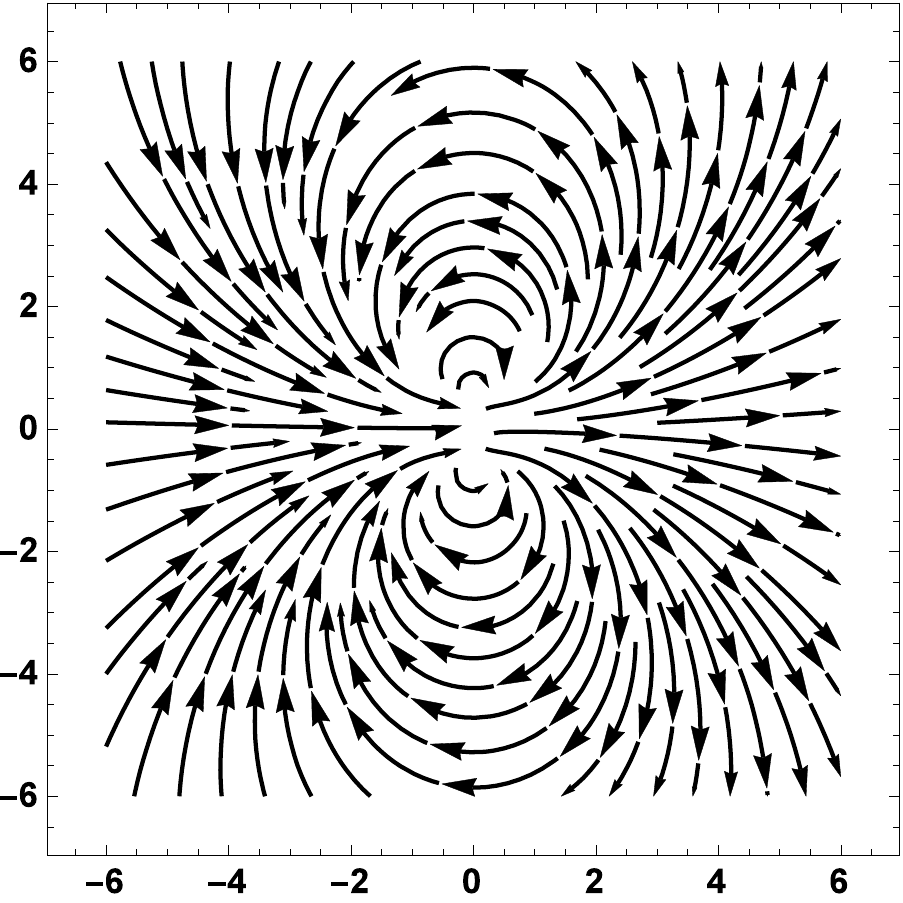}\\ (d) $\alpha=0$, $a=2$
	\label{7}
	\endminipage\hfill
\minipage{0.30\textwidth}
\includegraphics[width=2.0in,height=2.0in]{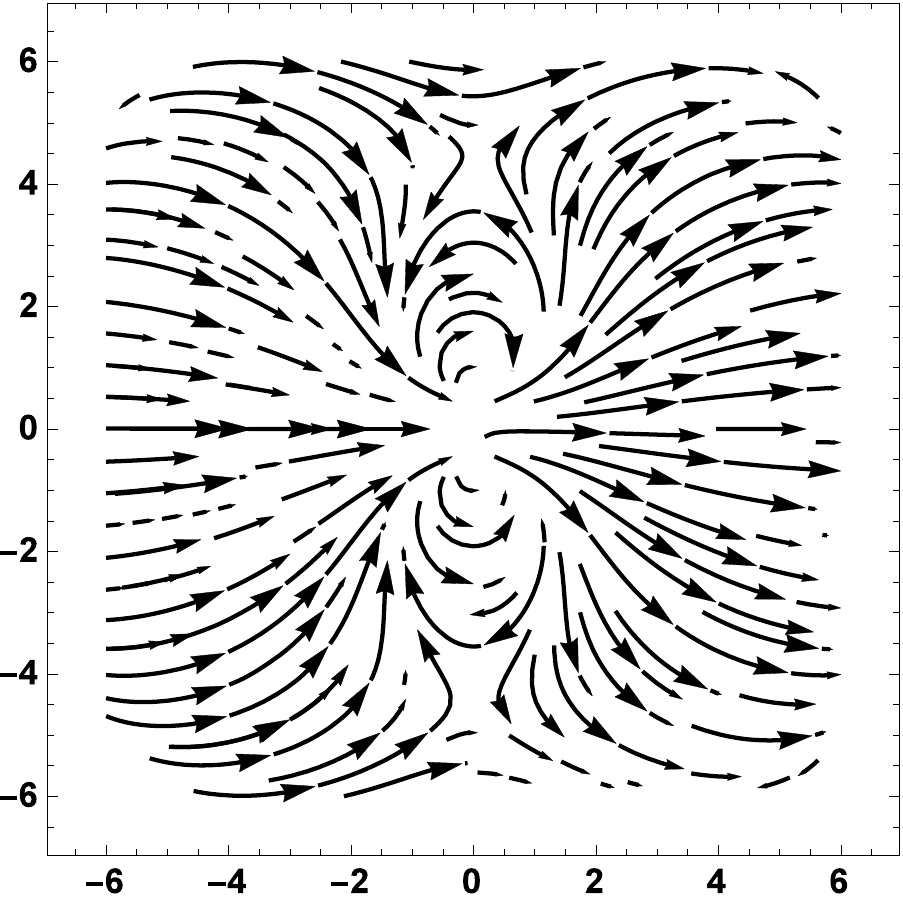} \\(e) $\alpha=.1$, $a=1.8$, $\omega_{q}=-2/3$
\label{7}
\endminipage\hfill
	\minipage{0.30\textwidth}
	\includegraphics[width=2.0in,height=2.0in]{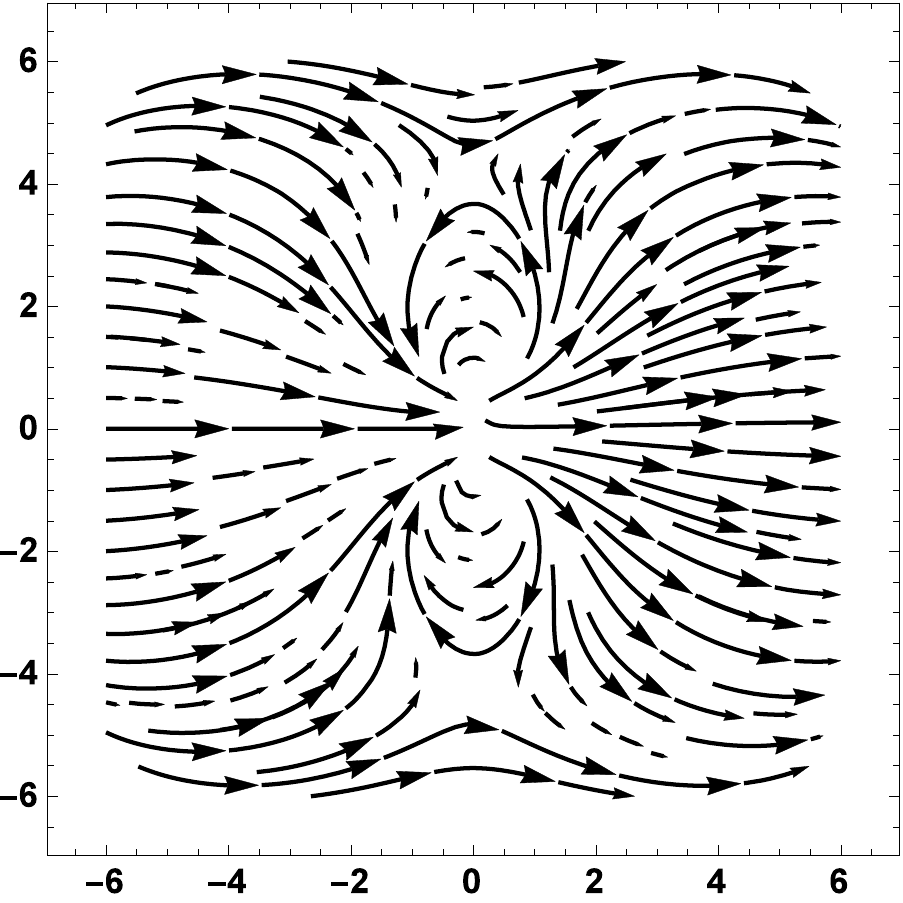}\\ (f) $\alpha=0.05$, $a=1.8$, $\omega_{q}=-7/9$
	\label{7}
	\endminipage\hfill
	\caption{{\footnotesize The vector field of the LT-~precession frequency \eqref{LT}, for black holes top (a-c) and for naked singularities bottom (d-f)  for different values of $\alpha$ (in Cartesian plane corresponding to $(r,\theta)$), is plotted in first and second row,  respectively. The field lines show that for black hole the vector field is defined outside the ergoshpere only, while for naked singularities it is finite up to the ring singularity.}}\label{VF}
\end{figure}
The magnitude of LT- precession frequency is
\begin{equation}
\Omega_{LT}=\frac{a\sqrt{ (2Mr+\alpha r^{b})^{2}%
		|\Delta| \cos ^{2}\theta +\left( M\left( r^{2}-a^{2}\cos ^{2}\theta \right) +%
		\frac{\alpha}{2}r^{b-1}\left\{ (2-b)r^{2}-ba^{2}\cos ^{2}\theta \right\} \right)
		^{2}\sin ^{2}\theta } }{\left( r^{2}+a^{2}\cos ^{2}\theta
	\right) ^{3/2}\left| r^{2}+a^{2}\cos ^{2}\theta -2Mr-\alpha r^{b}\right| }.
\end{equation}%
If we set $\alpha=0$, the LT- precession frequency for the Kerr black hole has been already  obtained  \cite{LTkerr}. The magnitude of the LT- precession frequency for different $\omega_{q}$ is plotted in FIG:\ref{LTmag} (a). The LT- precession frequency for black holes ($\omega_{q}=-4/9$) diverges at the ergosphere,  and for naked singularities ($\omega_{q}=-2/3$,~$\omega_{q}=-5/9$,~$\omega_{q}=-1/2$) it remains finite. Further, from FIG.\ref{LTmag} (b), we can see that for fixed $\alpha$, $\omega_{q}$ and $a$, for naked singularities  $\Omega_{LT}$ increases with increasing angle and has a peak. The peak increases with increasing angle,  and due to the ergosphere in the naked singularity case it blows up, as in the case of the Kerr black hole \cite{LTkerr}. 
\begin{figure}[!ht]
	\centering
	\minipage{0.50\textwidth}
	\includegraphics[width=8.8cm,height=6.0cm]{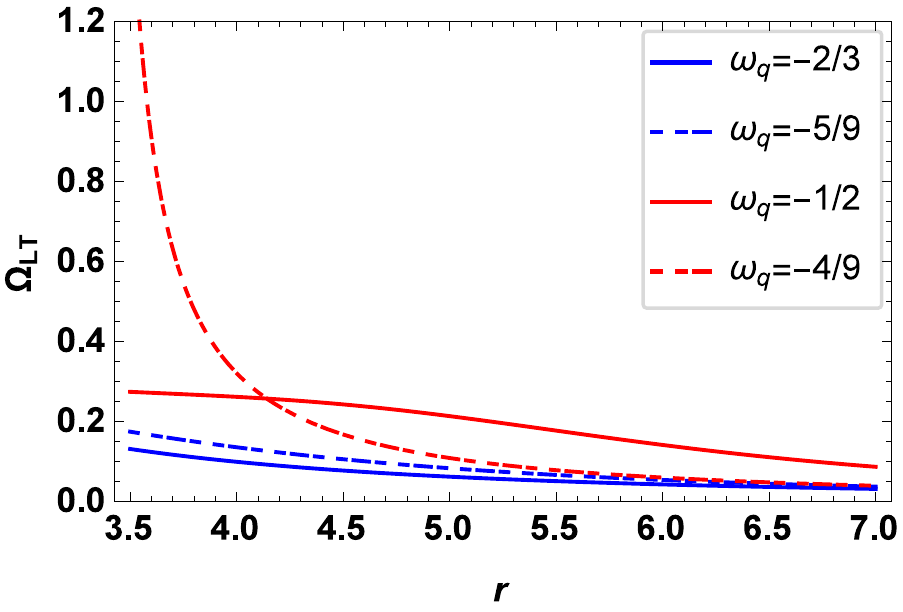}\\ (a) $\alpha=0.3$, $a=0.1$, $\theta=\pi/3$
	\label{6}
	\endminipage\hfill
	\minipage{0.50\textwidth}
	\includegraphics[width=8.8cm,height=6.0cm]{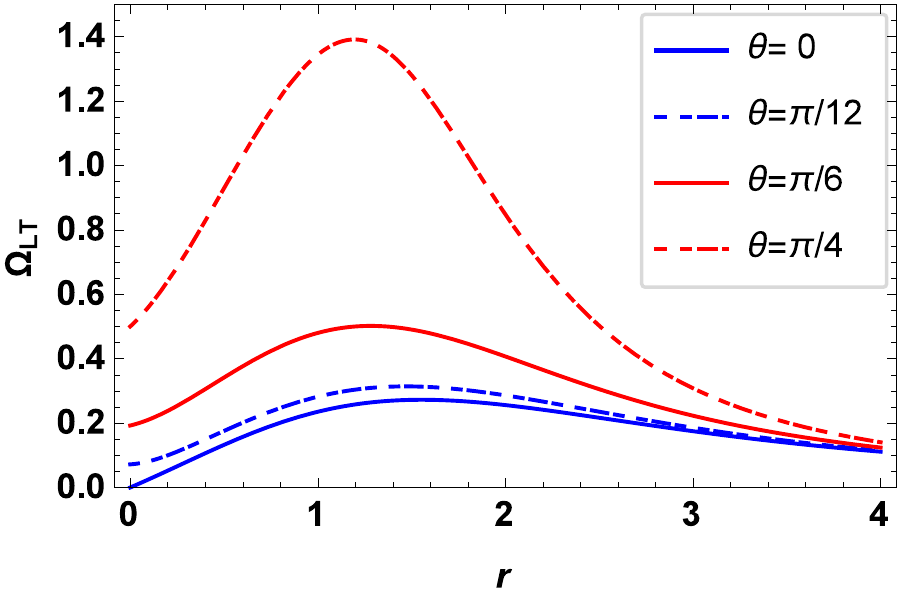}\\(b) $\alpha=0.2$, $a=2$, $\omega_{q}=-4/9$
	\label{7}
	\endminipage\hfill
	\caption{{\footnotesize (a) The magnitude of LT- precession frequency $\Omega_{LT}$ (in $M^{-1}$) versus $r$ (in $M$) against $r$ is plotted. The graph show that for fixed $\theta$, $\alpha$ and $a$, the LT- precession frequency remain finite for naked singularities with $\omega_{q}=-2/3,-5/9,-1/2$, while it blows up for black hole with $\omega_{q}=-4/9$ as the observer reach erfosphere. Further, for naked singularity with increasing $\omega_{q}$ the magnitude of $\Omega_{LT}$ increases. (b) The magnitude of $\Omega_{LT}$ for naked singularity is plotted which shows it is regular for throughout the region}}\label{LTmag}
\end{figure}

\subsection{Geodetic Precession}

If we set the spin parameter $a=0$ in the line element \eqref{BH}, it reduces to the KBH \cite{KBH}, in which the LT- precession frequency vanishes. However, for $a=0$ the precession frequency \eqref{SP} is non-zero. This precession is due to the curvature of the spacetime and known as the geodetic precession. It is given by
\begin{equation}\label{Oa=0}
\vec{\Omega}_{p}|_{a=0}=\Omega \frac{\left[ -\left( r^{2}-2Mr-\alpha r^{b}\right) \cos \theta \right]
\hat{r}+\left[ \left( r-3M+\frac{\alpha }{2}\left( b-4\right) r^{b-1}\right) \sin
	\theta \right] \hat{\theta} }{r-\left( 2M+\alpha r^{b-1}\right) -r^{3}\Omega
	^{2}\sin ^{2}\theta }.
\end{equation}%
As the KBH spacetime is spherically symmetric, the geodetic frequency is the same over the spherical surface around the black hole, so without loss of the generality we can set $\theta =\pi/2$. In the equatorial plane for any circular orbit of radius $r$ the angular frequency $\Omega$ of an observer is equal to its Kepler frequency $\Omega _{Kep}$, that is,  $\Omega=\Omega _{Kep}=\sqrt{ \frac{M}{r^{3}}+\frac{\alpha}{2}(2-b)r^{b-4}}$, and the magnitude of  \eqref{Oa=0} is given by
\begin{equation}
\Omega _{p}|_{a=0,\Omega =\Omega _{Kep}}=\Omega =\sqrt{\frac{%
		M}{r^{3}}+\frac{\alpha}{2}(2-b)r^{b-4}}.
\end{equation}%
The above expression is the precession frequency in the Copernican frame, computed with respect to the proper
time $\tau$. The proper time $\tau$, measured in the Copernican frame, is related to the coordinate time $t$ via 
\begin{equation*}
d\tau =\sqrt{ 1-\frac{3M}{r}+\frac{\alpha}{2}\left( b-4\right) r^{b-2}}~dt,
\end{equation*}
and we can obtain the geodetic precession frequency
in the coordinate basis as
\begin{equation*}
\Omega ^{^{\prime }}=\sqrt{\left( \frac{M}{r^{3}}+\frac{\alpha}{2}(2-b)r^{b-4}\right)
	\left( 1-\frac{3M}{r}+\frac{\alpha}{2}\left(b -4\right) r^{b-2}\right)}~,
\end{equation*}%
In terms of $\omega_{q}$, we get the geodetic precession frequency \footnote{It should be noted that the geodetic precession frequency obtained in \cite{Geodetic} has an error of a constant $2$.}\cite{Geodetic}
\begin{equation}
\Omega ^{^{\prime }}=\sqrt{\left( \frac{M}{r^{3}}+\frac{\alpha}{2}(1+3\omega_{q})r^{-3(1+\omega_{q})}\right)
	\left( 1-\frac{3M}{r}-\frac{3\alpha}{2}\left(1+\omega_{q}\right) r^{-(1+3\omega_{q})}\right)}.
\end{equation}
 Now, after a complete revolution of the observer around the black hole, the frequency associated with the change in the angle of the spin vector is given by
\begin{equation}\label{geo}
\Omega _{geodetic}=\sqrt{ \frac{M}{r^{3}}+\frac{\alpha}{2}(1+3\omega_{q})r^{-3(1+\omega_{q})}}\left( 1-\sqrt{ 1-\frac{3M}{r}-\frac{3\alpha}{2}\left( 1+\omega_{q}\right) r^{-(1+3\omega_{q})}}\right).
\end{equation}
For $\alpha=0$, we obtain the geodetic precession of the Schwarzschild black hole \cite{Geo,HartlelBook}. The geodetic precession frequency  is plotted in FIG. \ref{geodetic}, which shows that for fixed $\omega_{q}$, with increasing $\alpha$, the magnitude of the geodetic precession in a  circular orbit decreases (see FIG. \ref{geodetic} (a)), whereas for fixed $\alpha$ with increasing $\omega_{q}$ the magnitude increases. In addition,  for fixed $\omega_{q}$ and $\alpha$, the geodetic precession  decreases with increasing radius of the circular orbit (see FIG. \ref{geodetic}(b)).   

\begin{figure}[!ht]
	\centering
	\minipage{0.50\textwidth}
	\includegraphics[width=8.8cm,height=6.0cm]{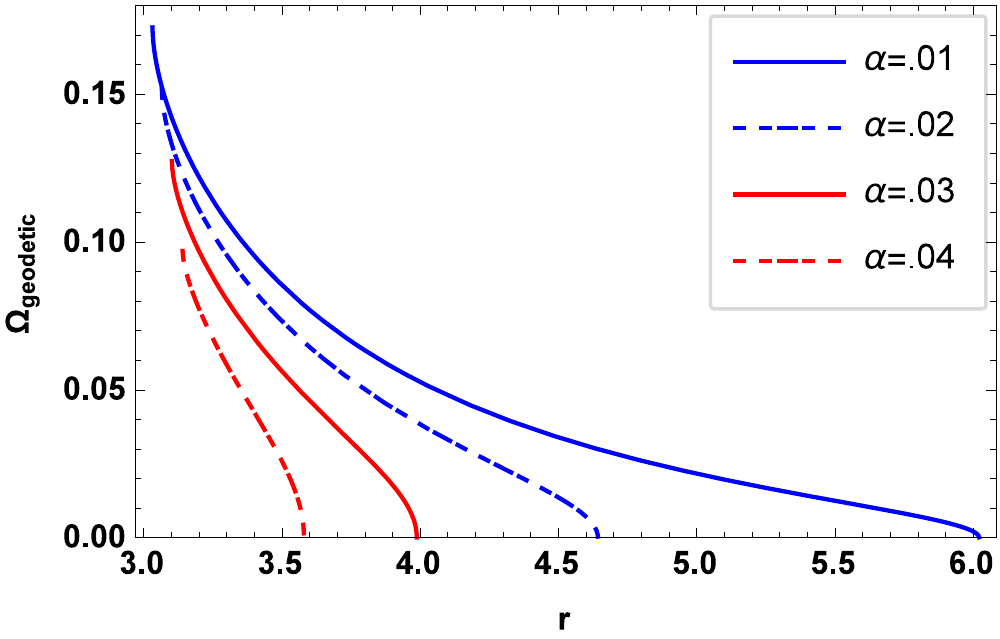}
	\label{6}
	\endminipage\hfill
	\minipage{0.50\textwidth}
	\includegraphics[width=8.8cm,height=6.0cm]{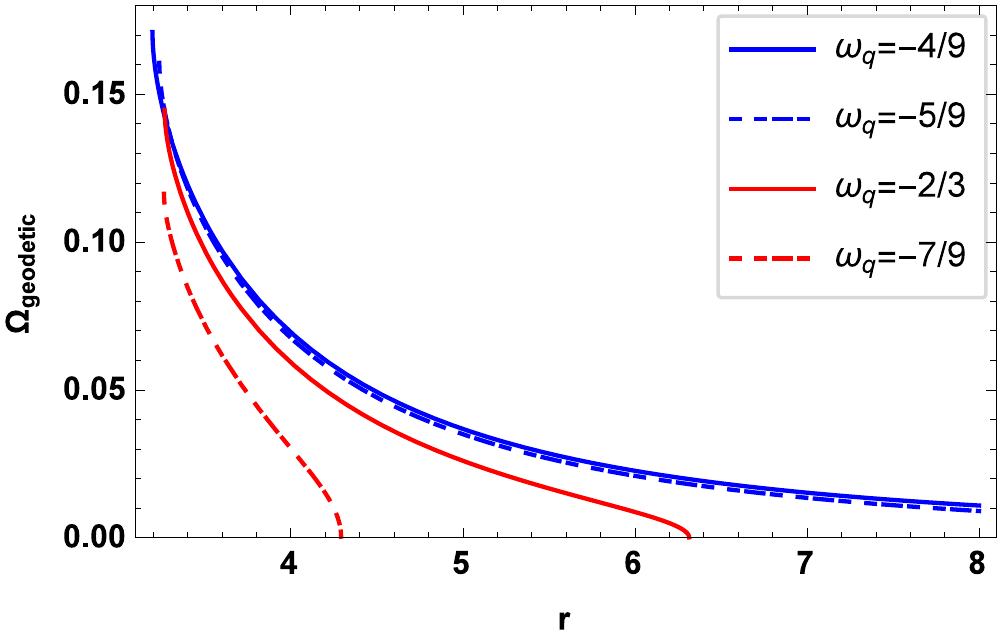}
	\label{7}
	\endminipage\hfill
	\caption{{\footnotesize (a) Geodetic precession $\Omega_{geodesic}$ versus $r$ is plotted for $M=1$ and $\omega_{q}=-8/9$. The graph shows that with increasing $\alpha$, the magnitude of $\Omega_{geodesic}$ in circular orbit decreases. (b) Geodetic precession against $r$ for $\alpha=0.05$ and different values of $\omega_{q}$ is plotted which shows that with increasing $\omega_{q}$,   $\Omega_{geodesic}$ in a fixed orbit increases. Further, for fixed $\omega_{q}$ and $\alpha$, geodetic precession decreases with increasing radius of circular orbit.}}\label{geodetic}
\end{figure}

\section{Distinguishing RKBHs from naked singularities using the precession of a test gyro }

In this section,  using the precession frequency of a test gyroscope attached to a stationary observer, we will differentiate RKBHs from naked singularities. The expression for the precession frequency is given in \eqref{SP}. For timelike stationary observers the angular velocity has a restricted range \eqref{omegarange}. The angular velocity $\Omega$ in terms of $\Omega_{\pm}$ can be written as
\begin{equation}\label{O}
\Omega =k\Omega _{+}+(1-k)\Omega _{-},
\end{equation}%
where $0<k<1$ and $\Omega_{\pm}$ defined by \eqref{AVm+-}. Using \eqref{omegarange} in \eqref{O} yields
\begin{equation}\label{AV}
\Omega =\frac{a\sin \theta \left( 2Mr+\alpha r^{b}\right) -\left( 1-2k\right)
	\Sigma \sqrt{\Delta }}{\sin \theta \left[ \left( r^{2}+a^{2}\right) \Sigma
	+a^{2}\sin ^{2}\theta \left( 2Mr+\alpha r^{b}\right) \right] }.
\end{equation}
For $k=1/2$, the angular velocity becomes
\begin{equation}\label{ozamo}
\Omega =\frac{a\sin \theta \left( 2Mr+\alpha r^{b}\right)}{\sin \theta \left[ \left( r^{2}+a^{2}\right) \Sigma
	+a^{2}\sin ^{2}\theta \left( 2Mr+\alpha r^{b}\right) \right] }=-\frac{g_{t\phi}}{g_{\phi \phi}}.
\end{equation}
The observer with this angular velocity is called the zero-angular-momentum observer (ZAMO). The precession frequency of the gyroscope attached to ZAMO in the Kerr black hole spacetime behaves different from a gyroscope attached to other observers having angular velocities different from ZAMO \cite{NS}. These gyros are non-rotating with respect to the local geometry and stationary observers regard both $+\phi$ and $-\phi$  \cite{zamo1,zamo2}. Thus, it is interesting to study how the precession frequency of the gyroscope attached to ZAMO behaves in the Kerr black hole surrounded by a quintessential matter field. Using \eqref{AV} in \eqref{SP}, we obtain the precession frequency in terms of the parameter $k$ as
\begin{equation}\label{opvector}
\vec{\Omega}_{p}=\frac{\left( r^{2}+a^{2}\right) \Sigma+a^{2}\sin
	^{2}\theta \left( 2Mr+\alpha r^{b}\right) }{4k\left( 1-k\right) \rho ^{7}\Delta }%
\left[ (F\sqrt{\Delta }\cos \theta) \hat{r}+(H\sin \theta)\hat{\theta}\right],
\end{equation}
where $F$ and $H$ are defined by \eqref{F&H}. Finally, the magnitude of precession frequency is given as
\begin{equation}\label{opkm}
\Omega_{p}=\frac{\left( r^{2}+a^{2}\right) \Sigma+a^{2}\sin
	^{2}\theta \left( 2Mr+\alpha r^{b}\right) }{4k\left( 1-k\right) \rho ^{7}|{\Delta}| }%
\left[ F^{2}|{\Delta }|\cos^{2}\theta+H^{2}\sin^2 \theta\right]^{1/2}.
\end{equation}
The denominator of the above equation vanishes at the ring singularity and horizons of the black hole. Also from \eqref{opvector}, we can see that the nominator of the radial part of ${\vec{\Omega}_{p}}$  goes to zero as the observer approaches  the horizons. So, we will study spin precessions when the observer reaches the horizon along different directions with different $k$, $a$, $\omega_{q}$ and $\alpha$ in detail. The magnitude of the precession frequency \eqref{opkm} versus $r$ is plotted for black holes in the left column and for naked singularities in the right column of FIG. \ref{opfig}, for $k=0.1,0.5,0.9$ in the first, second and third rows, respectively.

 For black holes with $a=0.7,~\alpha=0.1,~\omega_{q}=-5/9$, the precession frequency for $k=0.1$ and $k = ~0.9$ blows up, when the observer approaches the horizon along any given direction (see FIG. \ref{opfig}(a \& e)). On the other hand, in the case of naked singularities with ($a=1.3$,~$\omega_{q}=-5/9$,~$\alpha=0.1$), it can be seen from the right column of FIG. \ref{opfig}, the precession frequency for all $k=0.1,~0.5,~0.9$ remains finite up to $r=0$ along all directions except $\theta=\pi/2$. Near $r=0,~ \theta=\pi/2$, the frequency diverges because of the ring singularity, as in case of the Kerr black hole \cite{NS}. However, for ZAMO ($k=0.5$) the precession frequency in the RKBH behaves different from the Kerr black hole. For ZAMO, the precession frequency in the Kerr black hole remains finite as the observer approaches the horizon \cite{NS}, but for the RKBH it diverges along all directions except $\theta=\pi/2$ (see FIG.\ref{opfig}(c)). 
 \begin{figure}[!ht]
 	\centering
 	\minipage{0.50\textwidth}
 	\includegraphics[width=8.8cm,height=6.0cm]{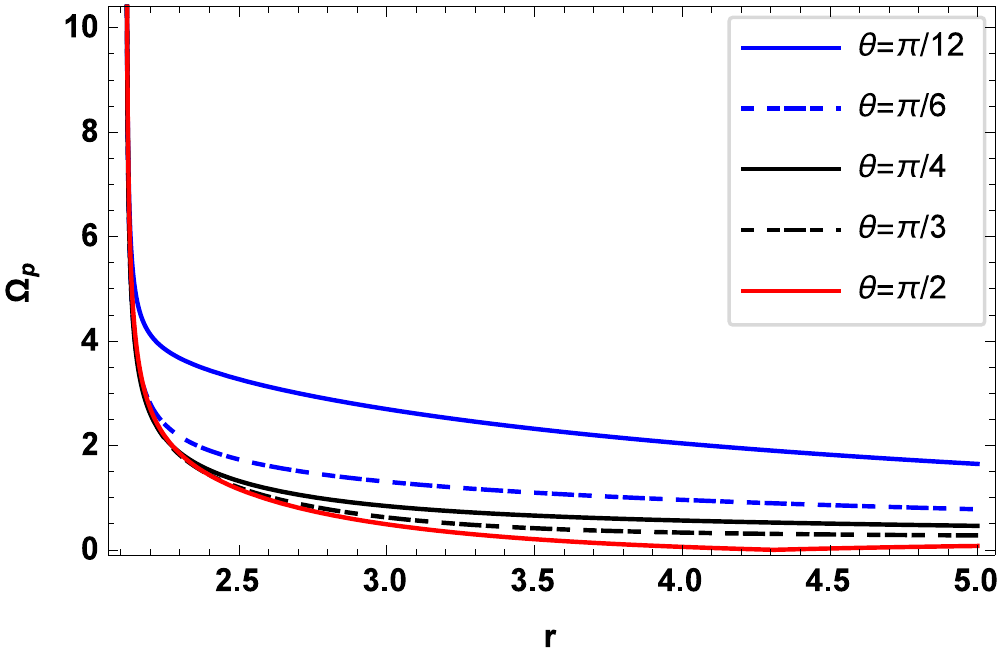}\\ (a) Black hole with $a=0.7$, $k=0.1$, $\alpha=0.1$, $\omega_{q}=-5/9$
 	\label{6}
 	\endminipage\hfill
 	\minipage{0.50\textwidth}
 	\includegraphics[width=8.8cm,height=6.0cm]{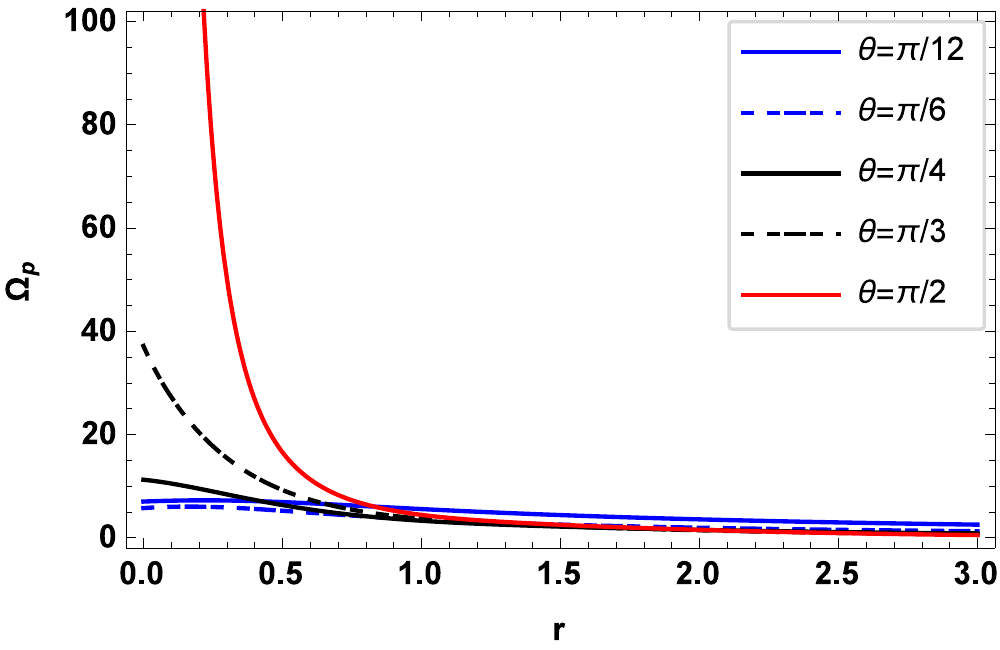}\\ (b) Naked singularity with $a=1.3$, $k=0.1$,$\alpha=0.1$, $\omega_{q}=-5/9$
 	\label{7}
 	\endminipage\hfill
 	\minipage{0.50\textwidth}
 	\includegraphics[width=8.8cm,height=6.0cm]{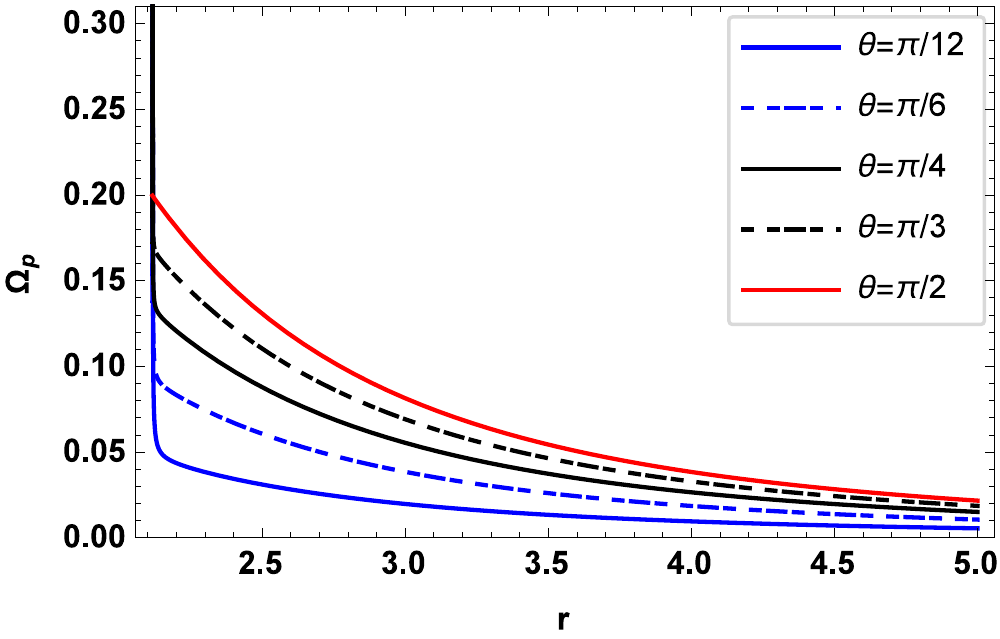}\\   (c) Black hole with $a=0.7$, $k=0.5$, $\alpha=0.1$, $\omega_{q}=-5/9$
 	\label{6}
 	\endminipage\hfill
 	\minipage{0.50\textwidth}
 	\includegraphics[width=8.8cm,height=6.0cm]{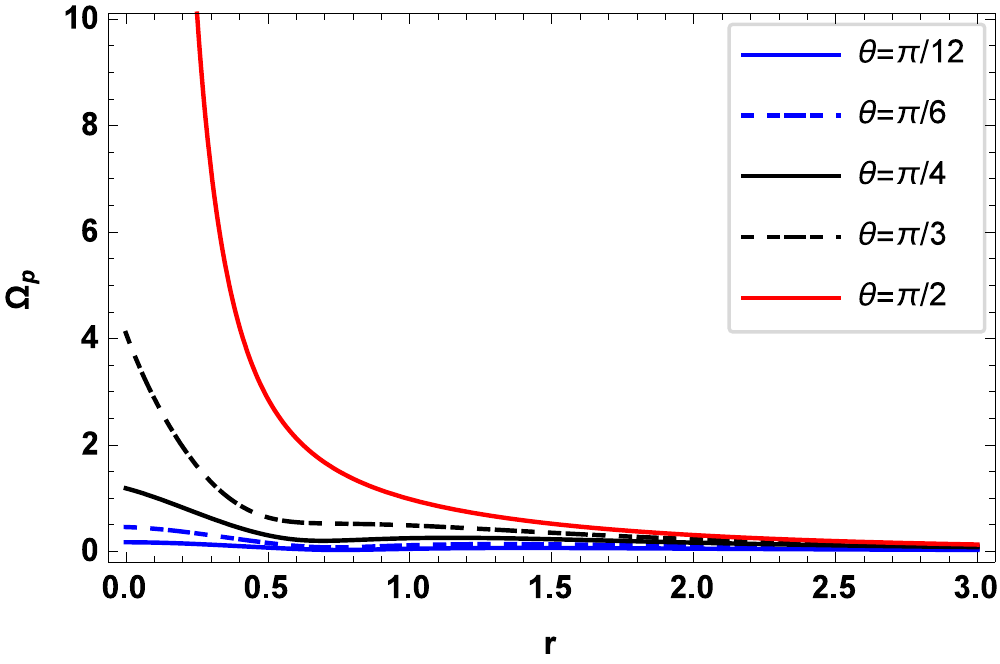}\\ (d) Naked Singularity with $a=1.3$, $k=0.5$, $\alpha=0.1$, $\omega_{q}=-5/9$
 	\label{6}
 	\endminipage\hfill
 	\minipage{0.50\textwidth}
 	\includegraphics[width=8.8cm,height=6.0cm]{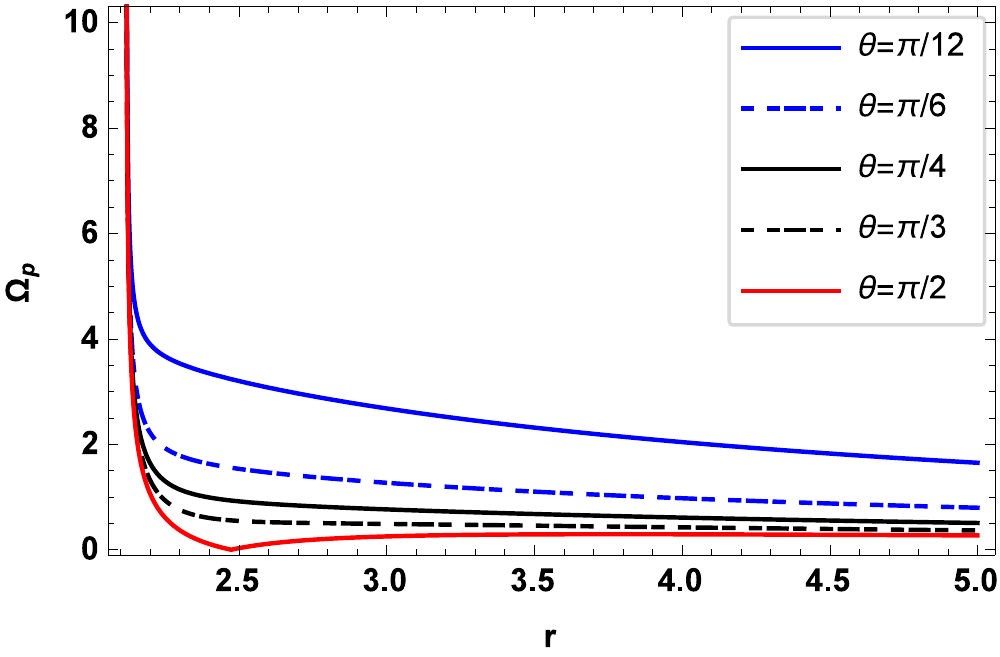}\\ (e) Black hole with $a=0.7$, $k=0.9$, $\alpha=0.1$, $\omega_{q}=-5/9$
 	\label{6}
 	\endminipage\hfill
 	\minipage{0.50\textwidth}
 	\includegraphics[width=8.8cm,height=6.0cm]{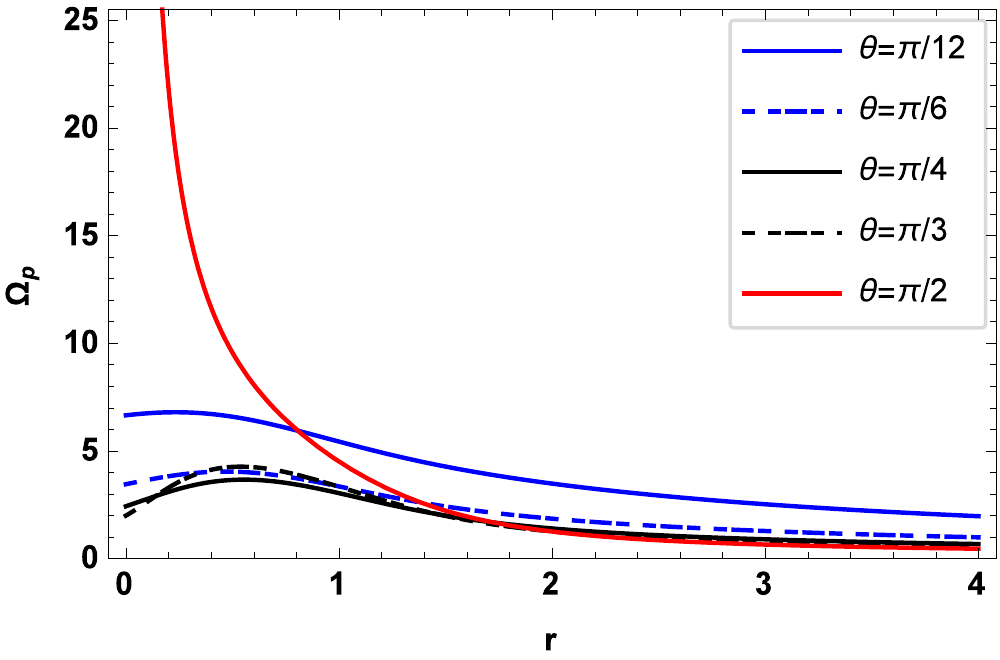}\\ (f) Naked Singularity with $a=1.3$, $k=0.9$, $\alpha=0.1$, $\omega_{q}=-5/9$
 	\label{7}
 	\endminipage\hfill
 	\caption{{\footnotesize It is plotted the magnitude of spin precession frequency $\Omega_p$ (in $M^{-1}$) versus $r$ (in $M$) for black holes in left column and for naked singularities in right column. For black holes we take $a=0.7$, $\alpha=0.1$, $\omega_{q}=-5/9$ and for naked singularities we take  $a=1.3$, $\alpha=0.1$, $\omega_{q}=-5/9$ and $k=0.1,0.5,0.9$ in first, second and third row, respectively. For black holes the precession frequency $\Omega_p$ diverges for all $k=0.1,0.5,0.9$ as the observer approaches the event horizon along any direction (except $\theta=\pi/2$ for $k=0.5$), whereas for naked singularities it remains finite along all directions except at the ring singularity ($r=0$,~$\theta=\pi/2$).}}\label{opfig}
 \end{figure}
\begin{figure}[!ht]
	\centering
	\captionsetup{justification=centering}
	\minipage{0.31\textwidth}
	\includegraphics[width=2.4in,height=1.7in]{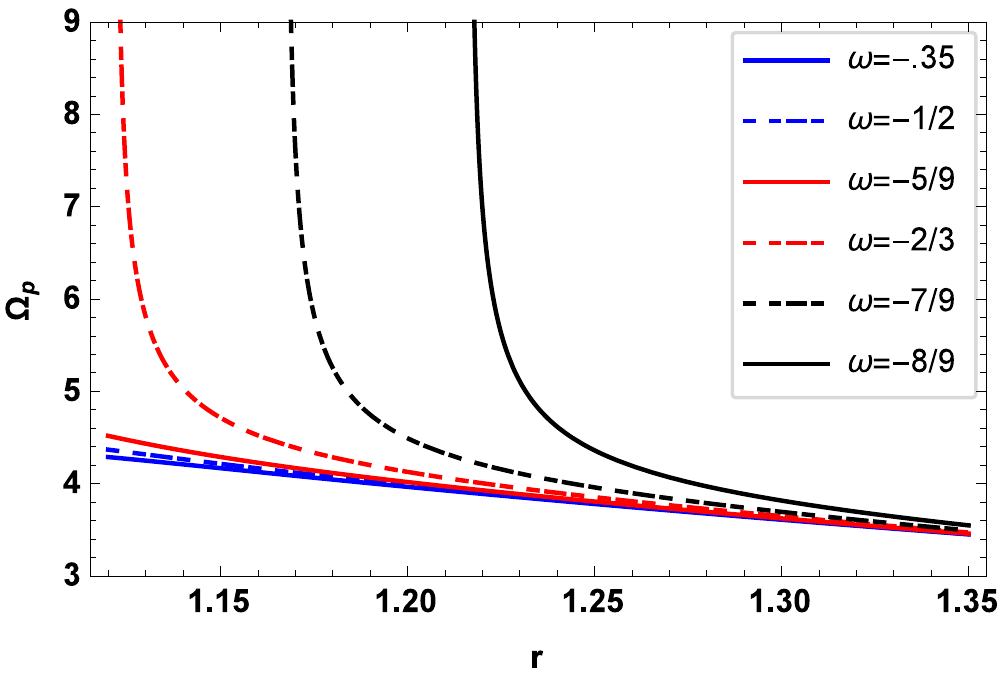}\\ (a)  $k=0.1$, $a=1.0275$, $\alpha=0.05$, $\theta=\pi/6$
	\label{6}
	\endminipage\hfill
	\minipage{0.31\textwidth}
	\includegraphics[width=2.4in,height=1.7in]{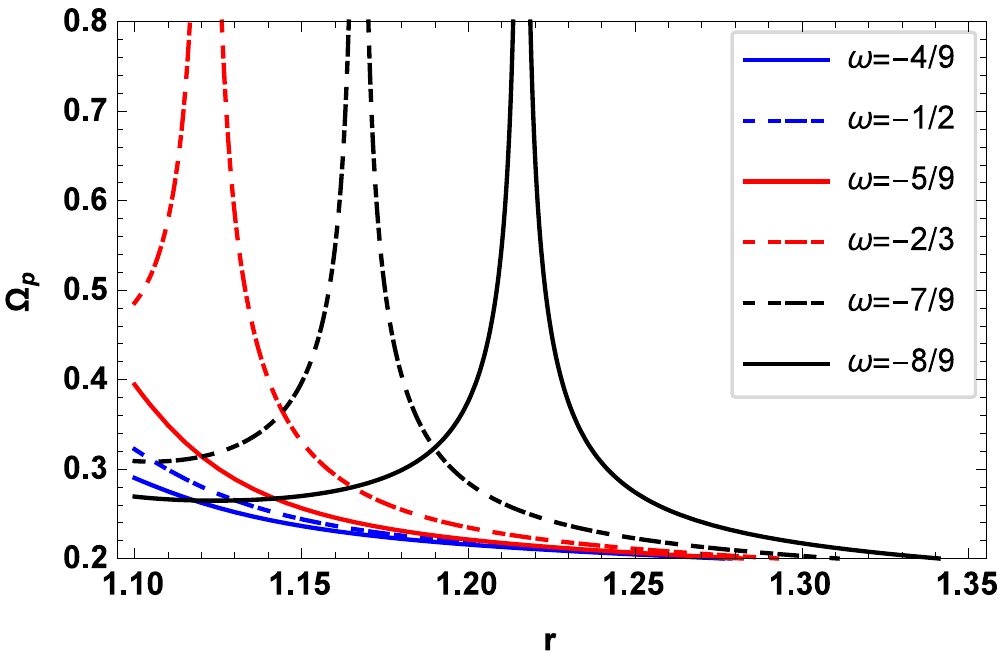}\\   (b)  $k=0.5$,  $a=1.0275$, $\alpha=0.05$, $\theta=\pi/6$
	\label{7}
	\endminipage\hfill
	\minipage{0.31\textwidth}
	\includegraphics[width=2.4in,height=1.7in]{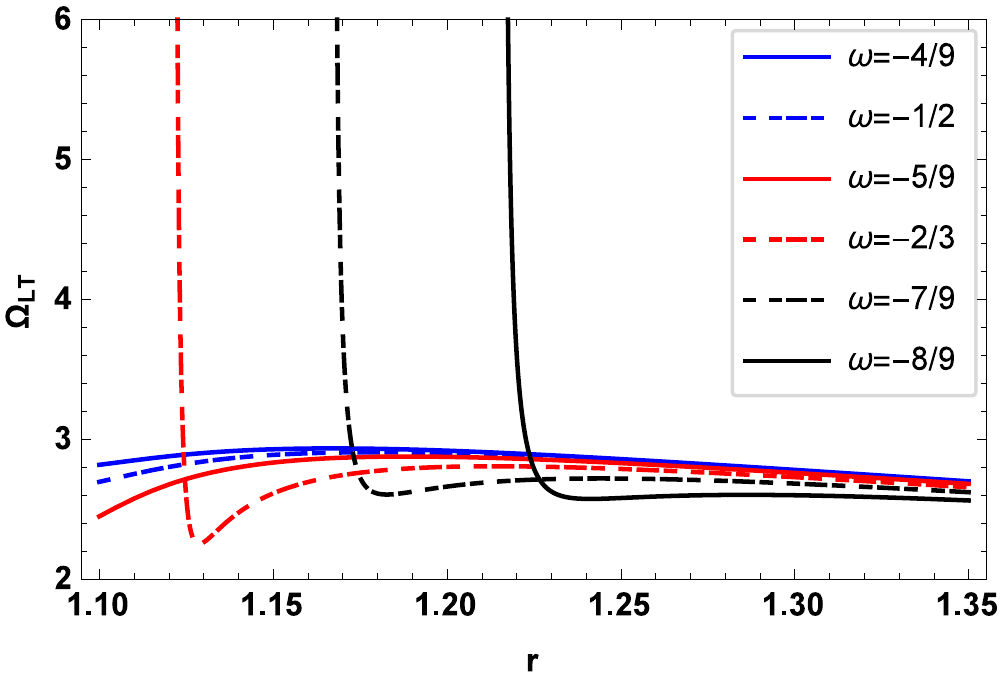}\\ (c) $k=0.9$, $a=1.0275$,  $\alpha=0.05$, $\theta=\pi/6$
	\label{7}
	\endminipage\hfill
	\minipage{0.31\textwidth}
	\includegraphics[width=2.4in,height=1.7in]{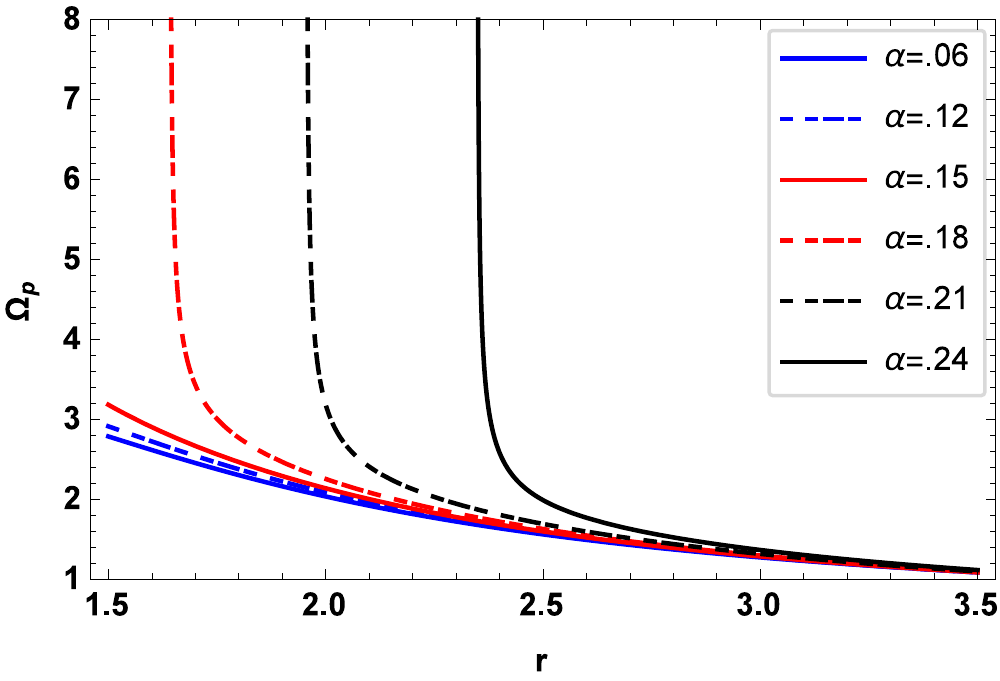}\\ (d) $k=0.1$, $\omega_{q}=-1/2$, $a=1.1$, $\theta=\pi/6$
	\label{6}
	\endminipage\hfill
	\minipage{0.31\textwidth}
	\includegraphics[width=2.4in,height=1.7in]{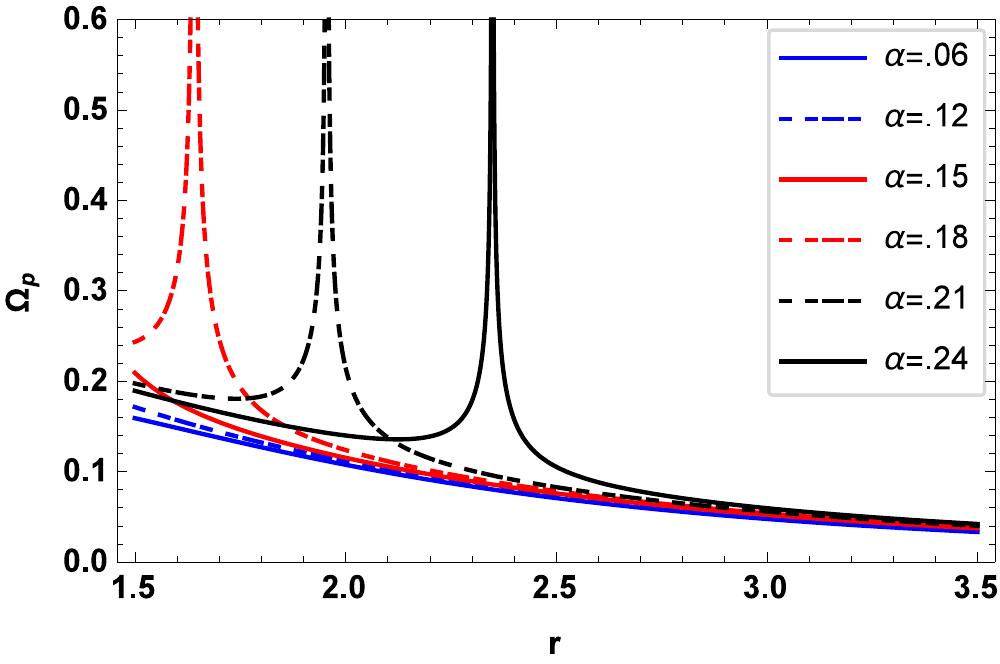}\\ (e) $k=0.5$, $\omega_{q}=-1/2$, $a=1.1$, $\theta=\pi/6$
	\label{7}
	\endminipage\hfill
	\minipage{0.31\textwidth}
	\includegraphics[width=2.4in,height=1.7in]{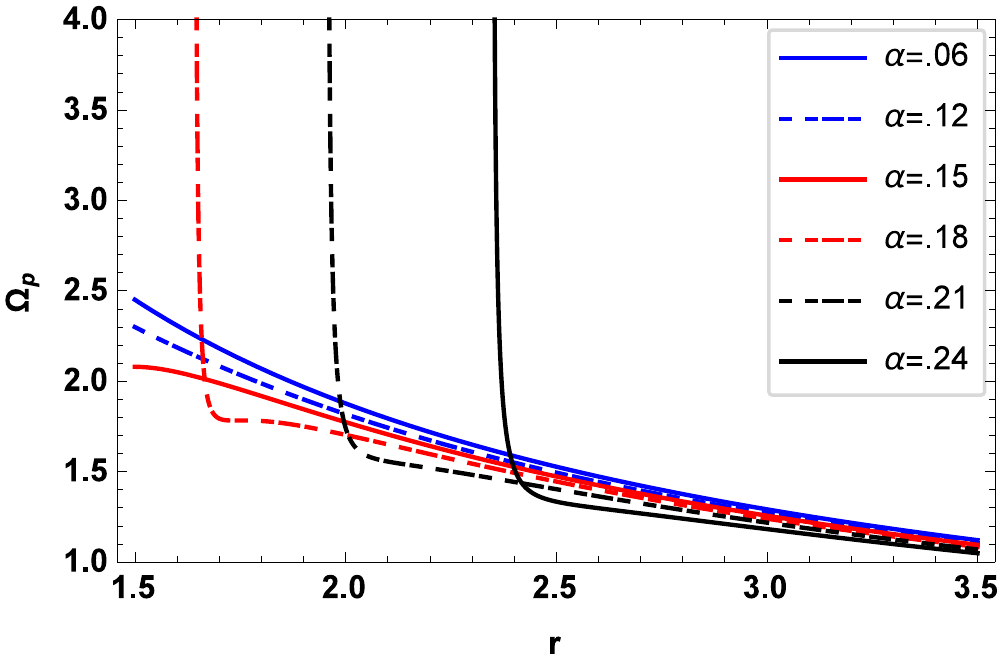}\\ (f) $k=0.9$, $\omega_{q}=-1/2$, $a=1.1$, $\theta=\pi/6$
	\label{7}
	\endminipage\hfill
	\minipage{0.31\textwidth}
	\includegraphics[width=2.4in,height=1.7in]{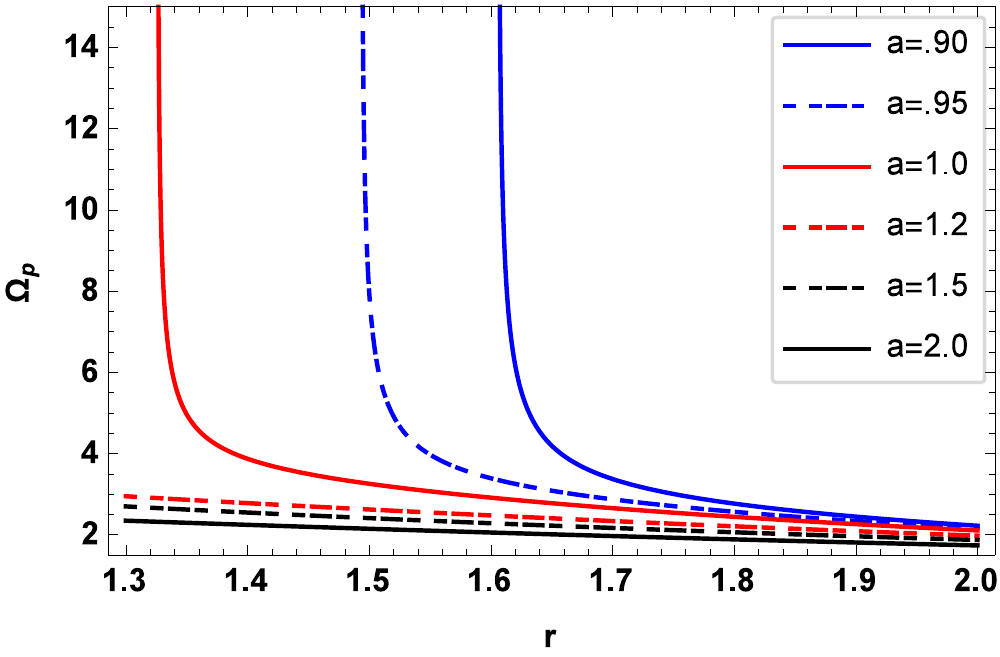}\\ (g) $k=0.1$,  $\omega_{q}=-5/9$,  $\alpha=0.05$, $\theta=\pi/6$
	\label{6}
	\endminipage\hfill
	\minipage{0.31\textwidth}
	\includegraphics[width=2.4in,height=1.7in]{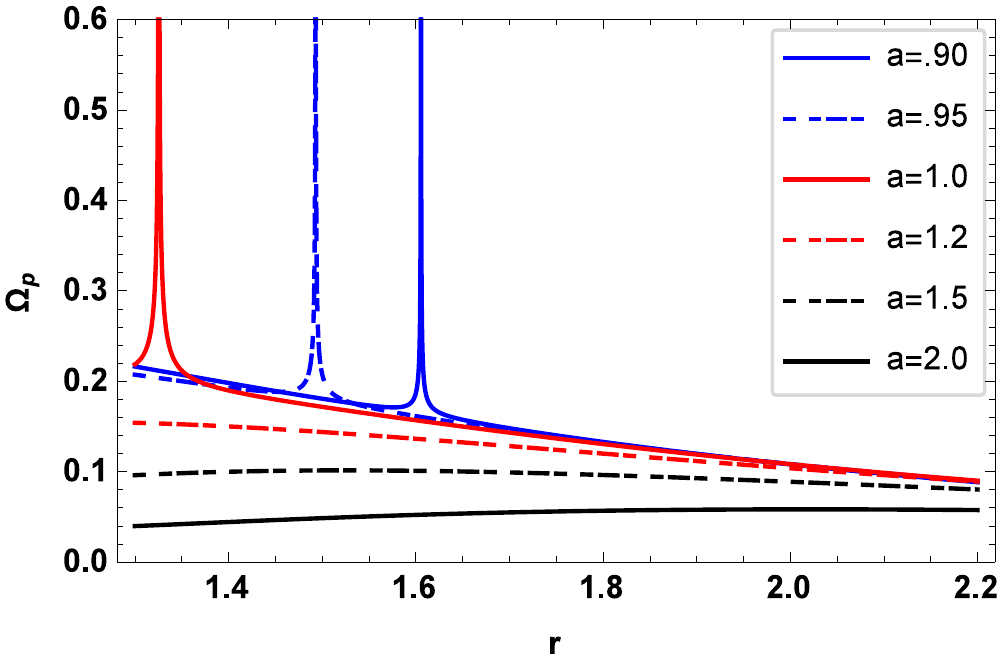}\\ (h) $k=0.5$, $\omega_{q}=-5/9$,  $\alpha=.05$, $\theta=\pi/6$
	\label{7}
	\endminipage\hfill
	\minipage{0.31\textwidth}
	\includegraphics[width=2.4in,height=1.7in]{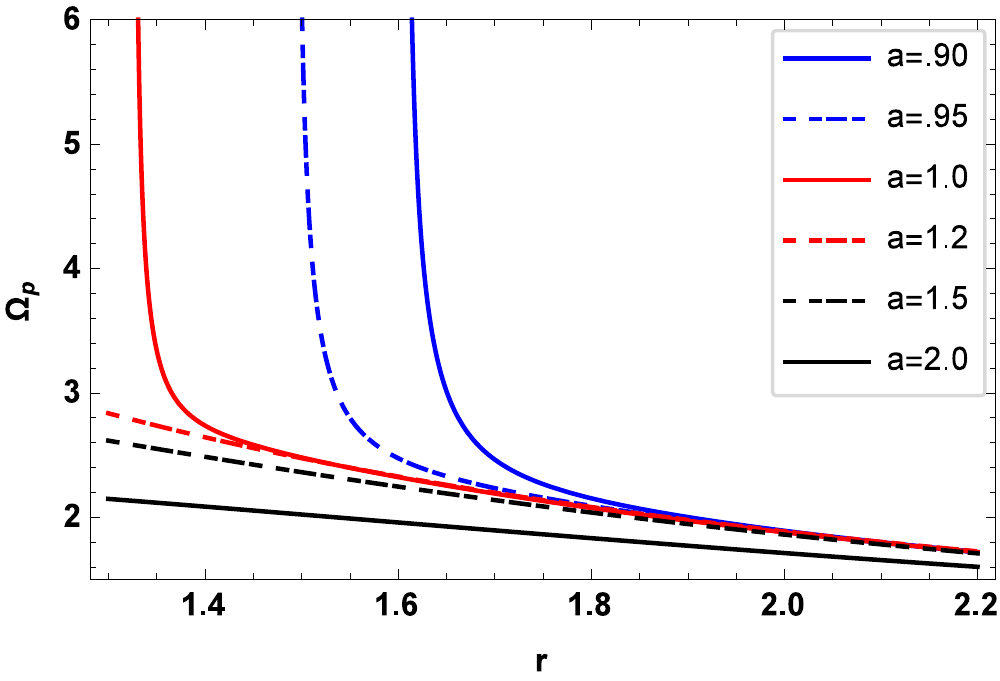}\\ (e) $k=0.9$,  $\omega_{q}=-5/9$,  $\alpha=0.05$, $\theta=\pi/6$
	\label{7}
	\endminipage\hfill
	\caption{{\footnotesize The variation of the magnetite of the precession frequency $\Omega_p$ (in $M^{-1}$) versus $r$ (in $M$) for different parameters is plotted. The graph shows that for black holes $\Omega_p$ diverges near the black hole horizon, while for naked singularities it remains finite. }}\label{Fig4}
\end{figure}

 For $k=0.5$, inserting \eqref{ozamo} into \eqref{F&H}, we get
\begin{eqnarray}
	F|_{k=0.5} &=&\frac{a^{3}\left( 2Mr+\alpha r^{b}\right) }{%
		8A^{2}}\left[ 8a\left( 2Mr+\alpha r^{b}\right) ^{2}\sin ^{4}\theta +A\left\{
	a^{2}-8Mr+4r^{2}+4\alpha r^{b}-a^{2}\cos 4\theta \right. \right. \nonumber\\ 
	&&\left. \left. +4\left( 2Mr-r^{2}+\alpha r^{b}\right) \cos 2\theta \right\} 
	\right], \label{F.5}\\  
	H|_{k=0.5} &=&\frac{-a\Delta \Sigma ^{2}}{4rA^{2}}\left[
	2Mr\left\{ a^{4}-3a^{2}r^{2}-6r^{4}+a^{2}\left( a^{2}-r^{2}\right) \cos
	2\theta \right\} \right. \nonumber \\ 
	&&\left. +\alpha r^{b}\left\{ ba^{4}+3\left( b-2\right) a^{2}r^{2}+2\left(
	b-4\right) r^{4}+a^{2}\left\{ a^{2}b+r^{2}\left( b-2\right) \right\} \cos
	2\theta \right\} \right], \label{H.5}
\end{eqnarray}%
where 
\begin{equation*}
A=\left( r^{2}+a^{2}\right) \Sigma +a^{2}\sin ^{2}\theta \left( 2Mr+\alpha
r^{b}\right), 
\end{equation*}
It is clear from \eqref{F.5} and \eqref{H.5} that, similarly to all other cases of $k$, for $k=0.5$ near the horizon the angular component of the precession frequency \eqref{opvector} remains finite but its radial component blows up. However, along $\theta=\pi/2$, the radial component is zero and thus along this direction near the horizon the precession frequency is finite. On the other hand, for the Kerr black hole, we have
\begin{eqnarray}
	F|_{k=0.5,~\alpha =0} &=&\frac{2Mra^{3}\Delta \Sigma
		^{2}\sin ^{2}\theta }{\left( r^{2}+a^{2})\Sigma +2Mra^{2}\sin ^{2}\theta
		\right) ^{2}}, \\ \label{F.5a=0}
	H|_{k=0.5,~\alpha =0} &=&-\frac{aM\Delta \Sigma ^{2}\left[
		a^{4}-3a^{2}r^{2}-6r^{4}+a^{2}\left( a^{2}-r^{2}\right) \cos 2\theta \right] 
	}{2\left( r^{2}+a^{2})\Sigma +2Mra^{2}\sin ^{2}\theta \right) ^{2}}. \label{H.5a=0}
\end{eqnarray}%
From \eqref{opkm}, we can see that in this case the precession frequency remains finite,  when the observer reaches the black hole along any given direction \cite{NS}. The gyroscope attached to all the stationary observers,  including ZAMO, in the KRBH behaves in the same manner, and the peculiar behavior of the ZAMO observers in the Kerr spacetime is avoided.

In FIG.\ref{Fig4}, we further illustrate the behavior of the precession frequency for other values of the parameter $a,~\omega_{q},~\alpha$. In the first, second and third rows of the figure, $\Omega_{p}$ versus $r$ is plotted for different values of $\omega_{q}$, $\alpha$ and $a$. It can be seen from the first row, for black holes ($\omega_{q}=-8/9,-7/9,-2/3$) and for all observers ($k=0.1,0.5,0.9$)  $\Omega_{p}$ diverges near the horizon. On the other hand, for naked singularities ($\omega_{q}=-7/20,-1/2,-5/9$), $\Omega_{p}$ and for all observers ($k=0.1,0.5,0.9$) it remains finite. In the second row, the behavior of $\Omega_{p}$ studied for different values of $\alpha$  shows that,  for all stationary observers ($k=0.1,0.5,0.9$) in the  black hole spacetimes  ($\alpha=0.18,0.21,0.24$),  $\Omega_{p}$ diverges near the horizon, but in the naked singularity spacetimes ($\alpha=0.06,0.12,0.15$),  it remains finite in the whole region. In can be also seen  that, the same behavior is also present for different values of $a$. That is, for all observers ($k=0.1,0.5,0.9$), the frequency diverges for black holes ($a=0.90,0.95,1.0$), whereas it remains finite for naked singularities. 

Finally, using the spin precession, we can differentiate the RKBHs from naked singularities. Consider a gyroscope attached to stationary observers with a nonzero azimuthal component ($\Omega$) of their four-velocity. These are observers moving along circles at constants $r$ and $\theta$, with a constant angular velocity $\Omega$. We can find the range of $\Omega$ such that their velocity is timelike. In this restricted range, we can define $\Omega$ in terms of the parameter $k$. Consider observers moving along two different directions $\theta_1$ and $\theta_2$. From the precession frequency $\Omega_p$ of the stationary observers moving in circular orbits we  conclude that: (i) if $\Omega_p$ becomes arbitrarily large as
approaching the central object in the quintessential field  along both $\theta_1$ and $\theta_2$, then the spacetime is a black hole. (ii) If $\Omega_p$ becomes arbitrary large when approaching to the central object for at most one of the two directions,
 the spacetime will be a naked singularity. For  black holes, $\Omega_p$ becomes arbitrarily large  when approaching  the event horizon, which covers the black hole singularity  in all directions, therefore observers approaching the black hole in all directions will see a divergence. However, for naked singularities, this divergence appears only along the ring singularly ($r=0,\theta=\pi/2$), therefore only observers along this direction will see the divergence. 
 
\section{Conclusions}

In this paper, we have presented critical values $\alpha_c$ and spin  $a_c$  of the quintessential  and spin parameters to distinguish the RKBHs from naked singularities. These values are directly proportional to the dimensionless parameter $\omega_{q}$, which has the range $-1<\omega_{q}<-1/3$. We have shown that,  if $\omega_{q}\rightarrow-1$, $\alpha_{c}\rightarrow2/27$, then black holes can formed for very small $\alpha$, and if $\omega_{q}\rightarrow-1/3$, $\alpha_{c}\rightarrow1$ with $a_c\rightarrow \infty$,  a highly spinning black hole can be formed. Further,  for all $-1<\omega_{q}<-1/3$, the black holes have three, inner, event and outer horizons. We have also studied extremal black holes and found the bounds of the horizons. For all $\omega_{q}$, by increasing $\alpha$ the size of the event horizon increases, while the size of the outer horizon decreases. We then studied the critical value of the quintessential parameter $\overline{a}_c$ for a KBH and found the radius of the extremal black holes. Similar to a RKBH,  in the case of a KBH, by increasing $\alpha$ the size of the event horizon increases whereas the size of the outer horizon decreases. 

We have also studied the spin precession frequency of a test gyroscope attached to a timelike stationary observer in the RKBH spacetime. For timelike stationary observers having angular velocity $\Omega$ with respect to a fixed star we have found the restricted ranges of $\Omega$. From the precession frequency for static observers ($\Omega=0$), we have obtained the LT-precession frequency. For a RKBH, the LT-precession frequency diverges as the observer approaches the ergosphere along any direction. On the other hand, for naked singularities it remains finite throughout the whole region except at the ring singularity. From the general precession frequency we than obtained the geodetic precession for observers in a KBH. The magnitude of the geodetic precession frequency in a  fixed circular orbit for a fixed $\omega_{q}$  decreases when increasing $\alpha$, whereas for a fixed $\alpha$ it increases with increasing $\omega_{q}$.

Using the spin precession frequency we have differentiated  black holes from naked singularities. The range of the angular velocity of a stationary observer can be parameterized by $k$. For $k=0.5$, the observer is called ZAMO. If the precession frequency of a test gyroscope attached to stationary observers moving along two different directions diverge as the observes approaches the horizon central object, then the spacetime is a black hole. If the precession frequency along most of the directions remains finite, then the spacetime is a naked singularity. This is because for black holes the precession frequency diverges as the observer approaches the horizon along all the directions, while for naked singularities this divergence appears only when the observer reaches the center of the spacetime along $\theta=\pi/2$.

\section*{Acknowledgements}

  This work is supported in part by National Natural Science Foundation of China (NNSFC), with the Grant Nos.: 11375153 (A.W.), 11675145(A. W.).

\end{document}